\documentclass[10pt,aps,prd,twocolumn,showpacs,superscriptaddress,longbibliography,nofootinbib]{revtex4-2}
\usepackage{mathrsfs,amsmath,amsthm,latexsym,amssymb,amsfonts,epsfig,cancel,enumerate,graphicx,txfonts,diagbox}
\usepackage[utf8]{inputenc}
\usepackage[T1]{fontenc}
\usepackage{lmodern}
\usepackage{multirow}
\usepackage[colorlinks=true,linkcolor=blue,citecolor=blue,urlcolor=blue]{hyperref}
\usepackage{orcidlink}
\usepackage{xcolor}
\usepackage{booktabs}
\usepackage{appendix}
\usepackage{mathtools}
\usepackage{placeins}
\definecolor{navy}{RGB}{0,0,150}

\allowdisplaybreaks
\newcommand{\RGU}{Department of Physics, The Assam Royal Global University, Guwahati-781035, Assam, India}
\newcommand{\UCC}{Programa de P\'os-Gradua\c c\~ao em F\'{\i}sica \& Coordena\c c\~ao do Curso de F\'{\i}sica -- Bacharelado, Universidade Federal do Maranh\~{a}o, 65085-580 S\~{a}o Lu\'{\i}s, Maranh\~{a}o, Brazil}

\begin{document}

\title{Thermodynamics and orbital structure of anti-de Sitter black holes in\\[1ex] Palatini-inspired nonlinear electrodynamics}

\author{Edilberto O. Silva\orcidlink{0000-0002-0297-5747}}
\email{edilberto.silva@ufma.br}
\affiliation{\UCC}

\author{Jo\~{a}o A. A. S.\ Reis\orcidlink{0000-0002-2831-5317}}
\email{joao.reis@uesb.edu.br}
\affiliation{Departamento de Ci\^{e}ncias Exatas e Naturais, \\ Universidade Estadual do Sudoeste da Bahia, Itapetinga (BA), 45700-000, Brazil}

\author{Faizuddin Ahmed\orcidlink{0000-0003-2196-9622}}
\email{faizuddinahmed15@gmail.com}
\affiliation{\RGU}

\date{\today}

\begin{abstract}
We construct a consistent anti-de Sitter completion of the static and spherically symmetric black-hole solution sourced by the Palatini-inspired nonlinear electrodynamics \(Y^n\) model. Starting from the Einstein--Hilbert action with a negative cosmological constant and the first-order PINLED sector, we derive the full set of field equations and show that the nonlinear electromagnetic solution preserves its original parametric structure, while the lapse function acquires the standard AdS contribution. We then analyze the horizon structure, Hawking temperature, extended phase-space thermodynamics, and the associated equation of state. In addition, we investigate null and timelike geodesics, with emphasis on the effective potentials, photon sphere, shadow radius for a static observer at finite distance, and innermost stable circular orbit. The resulting framework furnishes the exact AdS extension of the asymptotically flat PINLED black hole and provides a coherent basis for numerical and phenomenological studies of its thermodynamic, optical, and orbital properties.
\end{abstract}

\maketitle

\section{Introduction}\label{sec:1}

The theoretical description of electrically charged black holes within classical general relativity is encapsulated by the Reissner--Nordstr\"{o}m (RN) solution, which couples Einstein gravity to the linear Maxwell field. This framework, however elegant, suffers from two well-known pathologies: the divergence of the electromagnetic energy density at the origin and the persistence of a curvature singularity at $r=0$. Both difficulties motivate the search for physically motivated modifications of the standard electromagnetic sector that could either regularize the geometry or, more broadly, enrich the phenomenology of self-gravitating charged objects beyond the RN paradigm.

Nonlinear electrodynamics (NLE) provides the most natural arena for such modifications. Its roots trace back to the seminal proposals of Born and Infeld \cite{Born:1934gh}, who introduced an upper bound on the electric field strength to cure the divergence of the point-charge self-energy, and of Euler and Heisenberg \cite{Euler:1935zz}, who derived an effective NLE Lagrangian from one-loop quantum electrodynamics. Both models violate the superposition principle while reducing to Maxwell's theory in the weak-field limit. In the gravitational context, NLE sources have been shown to yield a rich variety of static, spherically symmetric black-hole solutions whose causal structure, thermodynamic behavior, and geodesic properties differ markedly from the RN case~\cite{Demianski:1986wx,AyonBeato:1998ub,AyonBeato:1999rg,AyonBeato:2000zs,Bronnikov:2000vy,Breton:2003tk,Kruglov:2015yua,Kruglov:2017mpj}. Of particular physical importance is the possibility, first demonstrated by Ay\'{o}n-Beato and Garc\'{\i}a~\cite{AyonBeato:1998ub}, of constructing regular black holes, geometries that are free of the central singularity, by coupling gravity to suitable NLE Lagrangians. The original regular black hole, proposed by Bardeen~\cite{Bardeen:1968}, was later given an NLE interpretation within this framework, establishing a systematic connection between singularity resolution and the nonlinear nature of the electromagnetic source~\cite{AyonBeato:2000zs,Bronnikov:2000vy}.

Recent years have witnessed a proliferation of NLE models motivated by requirements of causality, unitarity, and weak-energy-condition compliance, or by the desire to encode specific strong-field corrections in a classical effective action~\cite{Kruglov:2015yua,Kruglov:2017mpj,Balart:2014cga,Fan:2016hvf}. Among these, models formulated in a first-order, or Palatini-inspired, fashion have attracted growing interest. Drawing an analogy with the Palatini formulation of general relativity, in which the metric and the connection are varied independently, one can construct NLE theories in which the field strength $F_{\mu\nu}$ and the auxiliary tensor field $P_{\mu\nu}$ are treated as independent dynamical variables. This approach, termed Palatini-inspired nonlinear electrodynamics (PINLED) by Verbin \emph{et al.}~\cite{PhysRevD.111.084061}, yields a distinct class of field equations and a correspondingly distinct family of self-gravitating solutions. In the PINLED $Y^n$ model, defined by the first-order Lagrangian $L^{(1)}_{Y^n}$, where $Y = P^{\mu\nu}F_{\mu\nu}$, the static and spherically symmetric black-hole solutions and their optical and orbital properties in the asymptotically flat case have been studied in detail in Ref.~\cite{Cimdiker2026OpticalOrbital}.

The inclusion of a negative cosmological constant $\Lambda$ endows the problem with additional depth. Anti-de Sitter (AdS) spacetimes occupy a privileged role in theoretical physics, both as backgrounds for the AdS/CFT correspondence~\cite{Maldacena:1997re,Witten:1998qj,Gubser:1998bc} and as natural arenas for the study of black-hole thermodynamics. The thermodynamics of black holes in AdS were first systematically analyzed by Hawking and Page~\cite{Hawking:1982dh}, who identified a first-order phase transition between a large AdS-Schwarzschild black hole and a thermal AdS background. This transition was subsequently interpreted in the holographic context by Witten~\cite{Witten:1998zw} as the dual of a deconfinement transition in the boundary gauge theory. For charged black holes, Chamblin \emph{et al.}~\cite{Chamblin:1999tk} revealed a richer phase structure, including a first-order coexistence curve in the $(T, q)$ plane that closely mirrors the liquid-gas transition in classical fluids.

The extended phase-space formalism elevated this thermodynamic analogy to a more precise correspondence. Identifying the cosmological constant as a thermodynamic pressure, $P = -\Lambda/(8\pi G)$, with the black-hole mass playing the role of enthalpy rather than internal energy~\cite{Kastor:2009wy,Dolan:2010ha,Dolan:2011xt}, Kubiznak and Mann~\cite{Kubiznak:2012wp} demonstrated that the equation of state of charged AdS black holes formally reproduces the Van der Waals equation, complete with a critical point whose critical exponents coincide with those of the Van der Waals universality class. This framework, often called black-hole chemistry~\cite{Kubiznak:2016qmn}, has since been applied to a broad spectrum of solutions, uncovering phenomena such as reentrant phase transitions~\cite{Altamirano:2013ane}, triple points~\cite{Altamirano:2013uqa}, and superfluid-like transitions~\cite{Hennigar:2016xwd} that have no counterpart in the asymptotically flat setting. In the NLE sector, extended phase-space thermodynamics has been investigated for Born-Infeld AdS black holes~\cite{Gunasekaran:2012dq}, for exponential NLE~\cite{Kruglov:2017mpj}, and for a variety of other models~\cite{Hendi:2012um,Zangeneh:2015eza}, consistently revealing that the nonlinear coupling deforms the critical point and modifies the phase-transition structure relative to the RN-AdS baseline.

Independently of their thermodynamic properties, black-hole geometries are characterized by their geodesic structure. The unstable circular photon orbit, the photon sphere, governs both the size of the black-hole shadow~\cite{Synge:1966,Luminet:1979nyg,Bardeen:1973tla} and the ringdown spectrum of gravitational-wave perturbations in the eikonal limit~\cite{Ferrari:1984zz,Cardoso:2008bp}. For massive test particles, the innermost stable circular orbit (ISCO) is of direct astrophysical relevance, setting the inner edge of the accretion disk and, through the radiative efficiency, the luminosity of accreting systems. The landmark imaging of the supermassive black hole M87$^*$ by the Event Horizon Telescope (EHT)~\cite{EventHorizonTelescope:2019dse,EventHorizonTelescope:2019uob} and the subsequent imaging of Sgr~A$^*$~\cite{EventHorizonTelescope:2022wkp} have transformed the photon sphere and shadow into observational quantities. These measurements motivate detailed computations of shadow radii and photon-sphere positions for modified black-hole geometries to constrain alternative theories of gravity and exotic electromagnetic couplings. In the asymptotically AdS case, the natural framework for the shadow is that of a static observer at finite radius~\cite{Perlick:2015vta,Pantig:2022whj}, since the spacetime does not admit a globally defined spatial infinity in the standard sense.

In the present work, we construct the consistent anti-de Sitter extension of the PINLED $Y^n$ black hole by incorporating a negative cosmological constant directly into the Einstein-PINLED action. This variational approach guarantees that the nonlinear electromagnetic sector remains structurally unchanged, while the gravitational equation acquires the standard $\Lambda g_{\mu\nu}$ contribution. A central analytical result is that the mass equation governing the parametric solution is identical to that of the asymptotically flat case: the AdS geometry is generated solely by the replacement $f(\rho) \to f_{\rm flat}(\rho) + \rho^2/\tilde{L}^2$, where $\tilde{L}^2 = -3/\tilde{\Lambda}$ is the dimensionless AdS radius. Building on this foundation, we carry out a systematic study of the horizon structure, Hawking temperature, and extended phase-space thermodynamics, including the equation of state, the heat capacity at constant pressure, the Gibbs free energy, and the onset of Van der Waals-like phase transitions, as well as the null and timelike geodesic structure, with emphasis on the effective potentials, photon sphere, finite-distance shadow radius, and ISCO.

The manuscript is organized as follows. In Sec.~\ref{sec:2}, we present the Einstein--PINLED action with cosmological constant and derive the corresponding field equations. In Sec.~\ref{sec:3}, we obtain the static, spherically symmetric AdS black-hole solution in parametric form. In Sec.~\ref{sec:4}, we analyze the horizon structure and Hawking temperature, and present numerical results for representative parameter choices. In Sec.~\ref{sec:5}, we develop the extended phase-space thermodynamics and analyze the equation of state, heat capacity, Gibbs free energy, and critical behavior. In Sec.~\ref{sec:6}, we study null and timelike geodesics, computing the effective potentials, photon sphere, shadow radius for a finite-distance observer, and ISCO. Concluding remarks are given in Sec.~\ref{sec:7}. Throughout the paper, we work in geometrized units $G=c=1$; when discussing thermodynamic quantities, we additionally set $\hbar=k_B=1$. We adopt the metric signature $(+,-,-,-)$.

\section{Einstein--PINLED theory with cosmological constant}\label{sec:2}

\subsection{Motivation and action principle}

The standard formulation of nonlinear electrodynamics (NLE) coupled to gravity proceeds
in the second-order formalism: one specifies a Lagrangian density $\mathcal{L}(F)$ that
depends on the field-strength tensor $F_{\mu\nu}=\partial_\mu A_\nu-\partial_\nu A_\mu$ 
only through the Lorentz scalars $\mathcal{F}=\frac{1}{4}F_{\mu\nu}F^{\mu\nu}$ and
$\mathcal{G}=\frac{1}{4}F_{\mu\nu}{}^*\!F^{\mu\nu}$, and then varies the action with 
respect to the single dynamical field $A_\mu$. The resulting field equations are 
second-order in derivatives of $A_\mu$, and the solution space is in one-to-one 
correspondence with that of the Maxwell theory in appropriate limits.

The Palatini-inspired NLE (PINLED) approach, introduced in Ref.~\cite{PhysRevD.111.084061}, 
departs from this paradigm in an important way. Drawing an analogy with the 
Palatini (or first-order) formulation of general relativity, in which the metric 
$g_{\mu\nu}$ and the affine connection $\Gamma^\lambda{}_{\mu\nu}$ are treated as 
independent dynamical variables and varied separately, one promotes both the 
gauge potential $A_\mu$ and an auxiliary antisymmetric tensor field $P_{\mu\nu}$ 
to the status of independent dynamical variables in the electromagnetic action. 
The field strength $F_{\mu\nu}$ enters the Lagrangian as a shorthand for the curl 
of $A_\mu$, while $P_{\mu\nu}$ plays the role of a \emph{prepotential} or 
\emph{displacement-field tensor}. This structure is well known in the context of 
the $P$-representation of nonlinear electrodynamics~\cite{Plebanski:1970}, but the 
first-order variational principle gives rise to a fundamentally different class of 
theories: the Euler--Lagrange equations obtained from independent variation with 
respect to $A_\mu$ and $P_{\mu\nu}$ are not equivalent to those of any 
second-order NLE theory, and the solutions may therefore carry genuinely new 
physical features.

The complete action of the Einstein--PINLED theory with a negative cosmological 
constant is
\begin{equation}
S=\int d^4x\,\sqrt{-g}
\left[
\frac{1}{2\kappa}\left(R-2\Lambda\right)
+L^{(1)}_{Y^n}
\right],
\label{action}
\end{equation}
where $\kappa=8\pi G$, $\Lambda<0$ is the cosmological constant, and the first-order PINLED $Y^n$ Lagrangian density reads
\begin{equation}
L^{(1)}_{Y^n}
=
\frac{1}{4}P^{\mu\nu}P_{\mu\nu}
-\frac{1}{2}P^{\mu\nu}F_{\mu\nu}
+\frac{\gamma}{2n}\left(P^{\mu\nu}F_{\mu\nu}\right)^n
-J^\mu A_\mu.
\label{LYn}
\end{equation}
Here, $\gamma$ is a real coupling constant with dimensions of $[\text{field}]^{2(1-n)}$, $n>1$ is the nonlinearity index, and $J^\mu$ is an external current four-vector. The scalar $Y\equiv P^{\mu\nu}F_{\mu\nu}$ is the only independent Lorentz invariant that can be formed from the combination of $P_{\mu\nu}$ and $F_{\mu\nu}$ in the absence of magnetic sources.

It is instructive to examine the structure of $L^{(1)}_{Y^n}$ term by term. The quadratic piece $\frac{1}{4}P^{\mu\nu}P_{\mu\nu}$ is the kinetic term for the prepotential and plays a role analogous to the Legendre transform variable in the $P$-representation of electrodynamics. The term $-\frac{1}{2}P^{\mu\nu}F_{\mu\nu}
=-\frac{Y}{2}$ encodes the coupling between the auxiliary field $P_{\mu\nu}$ and the dynamical field strength $F_{\mu\nu}$; it is this mixed term that enforces the constitutive relation between $P_{\mu\nu}$ and $F_{\mu\nu}$ through the equations of motion. The nonlinear term $\frac{\gamma}{2n}Y^n$ introduces the 
PINLED self-interaction: for $n=1$ it merely renormalizes the kinetic term, so genuinely nonlinear dynamics require $n\geq 2$. Finally, the minimal coupling $-J^\mu A_\mu$ to an external current $J^\mu$ is included for generality and will be set to zero in the sourceless solutions of interest.

The dimensionless function
\begin{equation}
W(Y)=1-\gamma Y^{n-1}
\label{W_def}
\end{equation}
governs the deviation from linearity. For $\gamma=0$ (or $n=1$), $W=1$ and the theory reduces to the standard Maxwell theory. For $(-1)^n\gamma>0$ and $Y<0$ (the electric case, as we shall verify below), one has $W>1$, signaling an enhancement of the effective dielectric response relative to the vacuum Maxwell case. The sign and magnitude of $\gamma$ therefore determine the nature of the nonlinear electromagnetic medium encoded by the PINLED theory.

\subsection{Equations of motion}

Since $A_\mu$ and $P_{\mu\nu}$ are treated as independent fields, the action principle $\delta S=0$ is applied by varying them separately.

\paragraph{Variation with respect to $A_\mu$.}
Because $F_{\mu\nu}=\partial_\mu A_\nu-\partial_\nu A_\mu$ depends on $A_\mu$ through derivatives, integration by parts yields the generalized Amp\`{e}re--Gauss law
\begin{equation}
\nabla_\mu\!\left[W(Y)\,P^{\mu\nu}\right]=J^\nu.
\label{EOM_A}
\end{equation}
This is the \emph{source equation} of the PINLED theory. It generalizes the Maxwell equation $\nabla_\mu F^{\mu\nu}=J^\nu$ by replacing $F^{\mu\nu}$ with the $W$-dressed prepotential $W(Y)P^{\mu\nu}$. The factor $W(Y)$ in front of $P^{\mu\nu}$ plays the role of a field-dependent permittivity, and the source equation has the form of the macroscopic Maxwell equation $\nabla_\mu D^{\mu\nu}=J^\nu$ if one identifies the displacement field as $D^{\mu\nu}\equiv W(Y)P^{\mu\nu}$.

The Bianchi identity, which follows from the antisymmetry of $F_{\mu\nu}$ and is not affected by the first-order structure, reads
\begin{equation}
\nabla_{[\mu}F_{\nu\rho]}=0,
\label{Bianchi}
\end{equation}
as in standard electrodynamics.

\paragraph{Variation with respect to $P_{\mu\nu}$.} This yields the \emph{constitutive relation}
\begin{equation}
P_{\mu\nu}=W(Y)\,F_{\mu\nu}.
\label{constitutive}
\end{equation}
This equation, absent in the standard second-order formulation, is the hallmark of the Palatini-inspired approach: it relates the two independent tensor fields algebraically through the field-dependent factor $W(Y)$. Once Eq.~(\ref{constitutive}) is used to eliminate $P_{\mu\nu}$ in favor of $F_{\mu\nu}$, the scalar $Y$ becomes
\begin{equation}
Y
=P^{\mu\nu}F_{\mu\nu}
=W(Y)\,F^{\mu\nu}F_{\mu\nu}
=4W(Y)\,\mathcal{F},
\label{Y_self}
\end{equation}
which is a self-consistency equation for $Y$ in terms of the Maxwell invariant $\mathcal{F}$. The solution $Y=Y(\mathcal{F})$ of this equation maps the PINLED dynamics back to a (generically multi-valued) effective second-order theory. The key point is that for the $Y^n$ model, this mapping is \emph{not} the identity, so the PINLED $Y^n$ black-hole solutions are genuinely different from those of any standard NLE theory with the same field content.

In summary, the first-order PINLED field equations are
\begin{equation}
\nabla_\mu\!\left(W(Y)P^{\mu\nu}\right)=J^\nu,
\qquad
P_{\mu\nu}=W(Y)\,F_{\mu\nu},
\label{PINLED_eqs}
\end{equation}
where $W(Y)=1-\gamma Y^{n-1}$ and $Y=P^{\mu\nu}F_{\mu\nu}$.

\paragraph{Variation with respect to the metric.} Varying $\sqrt{-g}(R-2\Lambda)/(2\kappa)$ produces the Einstein tensor $G_{\mu\nu}$ plus the standard cosmological-constant term. Varying the matter Lagrangian 
$L^{(1)}_{Y^n}$ with respect to $g^{\mu\nu}$ and using the constitutive relation $P_{\mu\nu}=W(Y)F_{\mu\nu}$ to express everything in terms of $F_{\mu\nu}$, the total gravitational field equation reads
\begin{equation}
G_{\mu\nu}+\Lambda g_{\mu\nu}=-\kappa\,T_{\mu\nu},
\label{Einstein_AdS}
\end{equation}
where the electromagnetic energy-momentum tensor is
\begin{equation}
T_{\mu\nu}
=
-W^2(Y)\,F_{\mu\alpha}F^{\alpha}{}_{\nu}
-\frac{g_{\mu\nu}}{2}
\left[\frac{n-2}{2n}W(Y)-\frac{n-1}{n}\right]Y.
\label{Tmunu}
\end{equation}
Note that $T_{\mu\nu}$ is symmetric and traceless in the Maxwell limit $W\to 1$, $Y\to 4\mathcal{F}$, as required by the conformal invariance of Maxwell theory in four dimensions.

\subsection{Structure of the energy-momentum tensor and energy conditions}

It is useful to examine $T_{\mu\nu}$ in more detail. The first term, $-W^2 F_{\mu\alpha}F^\alpha{}_\nu$, is the standard electromagnetic stress tensor with a squared-$W$ prefactor. This term dominates in regions of strong field (large $|Y|$) and is responsible for the anisotropic electromagnetic pressure. The second term, proportional to $g_{\mu\nu}$, contributes an isotropic pressure that depends on both the nonlinearity index $n$ and the field strength through $Y$.

For a static, spherically symmetric, purely electric configuration (which is the case relevant to the present work), one has $Y\leq 0$ and $P^{\mu\nu}F_{\mu\nu}=-2\,P_{tr}F^{tr}<0$, so the scalar $Y$ is negative definite. The diagonal components of $T^\mu{}_\nu$ simplify to
\begin{align}
T^0{}_0=T^r{}_r
&=
-\frac{Y}{4}+\frac{3n-2}{4n}\,\gamma Y^n,
\\
T^\theta{}_\theta=T^\phi{}_\phi
&=
\frac{Y}{4}+\frac{n-2}{4n}\,\gamma Y^n.
\label{stress_components2}
\end{align}
The energy density measured by a static observer is $\rho_{\rm em}=-T^0{}_0$. Substituting $Y\leq 0$:
\begin{equation}
\rho_{\rm em}
=
\frac{|Y|}{4}+\frac{3n-2}{4n}\,(-1)^n\gamma|Y|^n.
\end{equation}
For $(-1)^n\gamma>0$ (which we assume throughout), both terms are positive and the weak energy condition $\rho_{\rm em}\geq 0$ is satisfied. This constraint on the sign of $\gamma$ is a necessary requirement for the physical viability of the solution and will be maintained in all subsequent analyses.

\subsection{Role of the cosmological constant}

A crucial observation is that $\Lambda$ appears \emph{exclusively} in the gravitational sector of the action~(\ref{action}) through the term $-\Lambda/\kappa$ in the Einstein--Hilbert Lagrangian. The PINLED Lagrangian 
$L^{(1)}_{Y^n}$ does not depend on $\Lambda$; hence, the electromagnetic field equations~(\ref{PINLED_eqs}) are completely unaffected by the presence of the cosmological constant. This is a direct consequence of the minimal coupling between gravity and matter: the cosmological term enters only through the modified gravitational equation~(\ref{Einstein_AdS}) via the replacement $G_{\mu\nu}\to G_{\mu\nu}+\Lambda g_{\mu\nu}$.

This observation has an important practical implication: the parametric solution for $F_{\mu\nu}$, $P_{\mu\nu}$, and $r(y)$ in the asymptotically flat PINLED geometry~\cite{PINLED:orbital:2025} carries over \emph{unchanged} to the AdS case. The only modification enters through the lapse function $f(\rho)$, which 
acquires the standard AdS correction $+\rho^2/\tilde L^2$ once the energy-momentum tensor~(\ref{Tmunu}) is substituted into the modified Einstein equation~(\ref{Einstein_AdS}). We therefore proceed to derive the AdS black-hole solution in Sec.~\ref{sec:3} using the same parametric strategy as in the asymptotically flat case, adapting only the gravitational sector.

\subsection{Maxwell limit and dimensional analysis}

Before closing this section, it is instructive to verify two consistency checks. First, in the \emph{Maxwell limit} $\gamma\to 0$, one has $W(Y)\to 1$ and the constitutive relation~(\ref{constitutive}) reduces to $P_{\mu\nu}=F_{\mu\nu}$. Substituting into Eq.~(\ref{EOM_A}) recovers the standard Maxwell equation $\nabla_\mu F^{\mu\nu}=J^\nu$, and the energy-momentum tensor~(\ref{Tmunu}) reduces to the Maxwell one, $T_{\mu\nu}=-F_{\mu\alpha}F^\alpha{}_\nu+\frac{1}{4}g_{\mu\nu}F_{\rho\sigma}F^{\rho\sigma}$. In this limit the full system reduces to the Reissner--Nordstr\"{o}m-AdS (RN-AdS) theory, providing the baseline against which the nonlinear corrections will be compared throughout the paper.

Second, the coupling constant $\gamma$ has dimension $[\gamma]=[\text{length}]^{2(n-1)}$ in geometrized units (i.e. $G=c=1$), so for $n=2$ it has dimension of length squared. The natural length scale associated with the PINLED sector is
\begin{equation}
\ell_{\rm PINLED}
\equiv
\left(|\gamma|E^2\right)^{-1/2}
\sim |\gamma|^{1/2},
\end{equation}
where $E$ is a characteristic electric field strength. In the dimensionless formulation introduced in Sec.~\ref{sec:3}, the fundamental length scale is $\ell=1/\sqrt{\kappa E^2}$, and the constraint $|\gamma|=E^{-2(n-1)}$ ensures that the nonlinear effects become comparable to the linear ones at the scale 
$\rho\sim\mathcal{O}(1)$, i.e.\ at $r\sim\ell$. For $r\gg\ell$, the Maxwell approximation is recovered, and the solution approaches the RN-AdS geometry. For $r\lesssim\ell$, the PINLED nonlinear corrections dominate and depart significantly from the Maxwell case.

\section{Static and spherically symmetric AdS solution}\label{sec:3}

\subsection{Ansatz and symmetry reduction}

We seek a static, spherically symmetric solution to the coupled Einstein-PINLED system derived in Sec.~\ref{sec:2}. The most general line element compatible with these symmetries takes the form \cite{Cimdiker2026OpticalOrbital}
\begin{equation}
ds^2
=
f(r)\,dt^2
-\frac{dr^2}{f(r)}
-r^2\!\left(d\theta^2+\sin^2\theta\,d\phi^2\right),
\label{metric}
\end{equation}
where $f(r)$ is the lapse function to be determined. The choice of a single metric function in the $(t,r)$ block is consistent with the absence of off-diagonal terms by Birkhoff's theorem for the spherically symmetric sector, and the equal-norm form ensures the simple relation $g_{tt}\,g_{rr}=-1$.

For the electromagnetic sector, the most general ansatz compatible with staticity and spherical symmetry allows only a radial electric field. Specifically, the non-zero independent components of $F_{\mu\nu}$ and $P_{\mu\nu}$ are
\begin{equation}
F_{tr}(r)=-F_{rt}(r)\neq 0,
\qquad
P_{tr}(r)=-P_{rt}(r)\neq 0,
\end{equation}
all other components vanish by symmetry. In this configuration, only a non-vanishing Maxwell scalar is
\begin{equation}
\mathcal{F}
=\tfrac{1}{4}F_{\mu\nu}F^{\mu\nu}
=\tfrac{1}{2}g^{tt}g^{rr}F_{tr}^2
=-\tfrac{1}{2f^2}F_{tr}^2\cdot f^2
=-\tfrac{1}{2}F_{tr}^2,
\end{equation}
which is negative for a purely electric field, so $\mathcal{F}<0$. The PINLED scalar $Y=P^{\mu\nu}F_{\mu\nu}$ likewise reduces to
\begin{equation}
Y
=2g^{tt}g^{rr}P_{tr}F_{tr}
=-\frac{2}{f(r)}\cdot f(r)\cdot P_{tr}F_{tr}
=-2P_{tr}F_{tr},
\end{equation}
and since $P_{tr}$ and $F_{tr}$ have the same sign (as we verify below from the constitutive relation), one has $Y\leq 0$ throughout, consistent with the electric nature of the field.

\subsection{Electromagnetic sector: parametric solution}

The key observation is that the electromagnetic field equations~(\ref{PINLED_eqs}) are independent of the cosmological constant (cf.\ Sec.~\ref{sec:2}). Therefore, the parametric solution for the fields $F_{tr}$, $P_{tr}$, and the implicit relation $r=r(Y)$ derived in the asymptotically flat case~\cite{PhysRevD.111.084061,PINLED:orbital:2025} is preserved unchanged in the AdS background.

To derive this solution explicitly, we use the generalized source equation $\nabla_\mu(W(Y)P^{\mu\nu})=0$ (sourceless case). For the radial component $\nu=t$, this yields
\begin{equation}
\partial_r\!\left(\sqrt{-g}\,W(Y)\,P^{rt}\right)=0,
\end{equation}
where $\sqrt{-g}=r^2\sin\theta$. Integrating with $P^{rt}=-g^{rr}g^{tt}P_{rt}
=P_{tr}/f\cdot f = P_{tr}$ (after raising indices with the metric), the conserved quantity is identified as the electric charge $Q$:
\begin{equation}
\sqrt{-g}\,W(Y)\,P^{rt}=Q
\implies
r^2\,W(Y)\,P_{tr}=Q.
\label{charge_int}
\end{equation}
Simultaneously, the constitutive relation $P_{\mu\nu}=W(Y)F_{\mu\nu}$ gives $P_{tr}=W(Y)F_{tr}$, so that $W^2(Y)F_{tr}=Q/r^2$.

Solving these two equations for $F_{tr}$ and $P_{tr}$ in terms of $Y$, which satisfies $Y=-2P_{tr}F_{tr}=-2W(Y)F_{tr}^2$, one obtains the parametric representation
\begin{align}
&F_{tr}=\sqrt{\frac{-Y}{2W(Y)}},
\\
&P_{tr}=\sqrt{\frac{-W(Y)\,Y}{2}},
\\
&r=\left(\frac{2Q^2}{-Y\,W^3(Y)}\right)^{1/4}.
\label{parametric_fields}
\end{align}
Here $Y$ serves as the parametric variable running over $(-\infty,0]$, with $Y\to 0$ corresponding to $r\to\infty$ (weak-field region) and $|Y|\to\infty$ corresponding to $r\to 0$ (strong-field region near the origin). For $Y\leq 0$ and $(-1)^n\gamma>0$ one verifies that $W(Y)=1-\gamma Y^{n-1}>0$, so both $F_{tr}$ and $P_{tr}$ are real and positive, confirming the self-consistency of 
the electric ansatz.

The third relation in Eq.~(\ref{parametric_fields}) provides the crucial \emph{implicit radial equation}: rather than writing the fields as explicit functions of $r$, the PINLED theory naturally admits a parametric representation in which $r$ and all field components are expressed as functions of $Y$ (or equivalently of the dimensionless variable $y=-Y/E^2$ introduced below). This parametric structure is a distinctive feature of the PINLED $Y^n$ model; it arises because the constitutive relation introduces a field-dependent dressing factor $W(Y)$ that makes the inversion $r\to Y(r)$ non-trivial in closed form.

The corresponding stress-energy tensor components, obtained by substituting the parametric solution into Eq.~(\ref{Tmunu}), take the diagonal form
\begin{align}
T^0{}_0=T^r{}_r
&=
-\frac{Y}{4}+\frac{3n-2}{4n}\,\gamma Y^n,
\label{Trr}\\
T^\theta{}_\theta=T^\phi{}_\phi
&=
\frac{Y}{4}+\frac{n-2}{4n}\,\gamma Y^n.
\label{Tthth}
\end{align}
The equality $T^0{}_0=T^r{}_r$ reflects the radial pressure isotropy of a purely electric, spherically symmetric configuration, analogous to the well-known relation $\rho=p_r$ for standard Maxwell electrodynamics. The tangential components $T^\theta{}_\theta$ differ from the radial ones due to the nonlinear self-interaction encoded in the $\gamma Y^n$ term. For $(-1)^n\gamma>0$ and $Y\leq 0$, the energy density $\rho_{\rm em}=-T^0{}_0=|Y|/4+[(3n-2)/(4n)] (-1)^n|\gamma||Y|^n>0$, confirming that the weak energy condition is satisfied.

\subsection{Dimensionless formulation}

The dimensional scales present in the problem, the gravitational constant $G=\kappa/(8\pi)$, the PINLED coupling $\gamma$, the electric charge $Q$, and the characteristic field strength $E$, can be combined into a single fundamental length
\begin{equation}
\ell
=\frac{1}{\sqrt{\kappa E^2}},
\label{ell_def}
\end{equation}
where $E$ is an arbitrary reference electric field with the dimension of the inverse of length squared in geometrized units ($[E]=\mathrm{length}^{-2}$). The PINLED coupling constant is then fixed by requiring that the nonlinear corrections become of order unity at the scale $r\sim\ell$:
\begin{equation}
|\gamma|=\frac{1}{E^{2(n-1)}},
\label{gamma_fixed}
\end{equation}
so that $\gamma E^{2(n-1)}=\pm 1$ and the nonlinearity parameter is absorbed into the field scale $E$. This is a natural normalization condition: it ensures that the dimensionless PINLED contribution $\gamma Y^{n-1}$ evaluated at $Y\sim -E^2$ is of order unity, so that the nonlinear corrections are important precisely 
in the strong-field region $r\lesssim\ell$.

We introduce the following set of dimensionless variables:
\begin{align}
&\rho=\frac{r}{\ell},
\qquad
m(\rho)=\frac{M(r)}{\ell},
\qquad
q=\kappa\,E\,Q,
\\
&y=-\frac{Y}{E^2},
\qquad
\tilde\Lambda=\Lambda\ell^2.
\label{dimless_vars}
\end{align}
Here $\rho$ is the dimensionless radial coordinate, $m(\rho)$ is the dimensionless Misner--Sharp mass function, $q$ is the dimensionless charge parameter, $y>0$ is the dimensionless field invariant (note the sign convention: $y=-Y/E^2>0$ since $Y\leq 0$ for electric fields), and $\tilde\Lambda<0$ is the dimensionless cosmological constant. In terms of these variables, the metric retains the same 
functional form~(\ref{metric}) with $r\to\rho\ell$, and the dimensionless AdS radius is $\tilde L^2=-3/\tilde\Lambda>0$.

The dimensionless energy density and radial relation become
\begin{equation}
K(y)
=\frac{y}{4}+\frac{3n-2}{4n}\,y^n,
\qquad
\rho(y)
=\left(\frac{2q^2}{y\,(1+y^{n-1})^3}\right)^{1/4},
\label{Ky_rho}
\end{equation}
where we used $W(Y)=1-\gamma Y^{n-1}=1+(-1)^n(-E^2 y)^{n-1}/E^{2(n-1)} =1+y^{n-1}$ (for $(-1)^n\gamma>0$). Thus $W(y)=1+y^{n-1}$ in dimensionless variables, and the condition $W>0$ is automatically satisfied for all $y\geq 0$.

\subsection{Einstein equations in the AdS background}

In terms of the dimensionless variables, the $(tt)$ component of the Einstein equation~(\ref{Einstein_AdS}) can be written as a first-order ordinary differential equation for the lapse function. For the static spherically symmetric metric~(\ref{metric}), the Einstein tensor components reduce to
\begin{equation}
G^t{}_t = G^r{}_r = \frac{1}{r^2}\frac{d}{dr}\!\left[r(1-f)\right]
-\frac{2}{r}\frac{df}{dr},
\end{equation}
and after combining with the cosmological term $\Lambda g^t{}_t=-\Lambda$, the field equation $G^\mu{}_\nu+\Lambda\delta^\mu{}_\nu=-\kappa T^\mu{}_\nu$ in the dimensionless radial coordinate takes the compact form
\begin{equation}
\frac{1}{\rho^2}\frac{d}{d\rho}\!\left[\rho(1-f)\right]
=K(\rho)+\tilde\Lambda.
\label{Einstein_first_order_AdS}
\end{equation}
Here $K(\rho)\equiv K(y(\rho))$ is the dimensionless energy density evaluated along the parametric curve $\rho=\rho(y)$, and the $\tilde\Lambda$ contribution comes directly from the $\Lambda g_{\mu\nu}$ term in Eq.~(\ref{Einstein_AdS}).

Comparing with the asymptotically flat case ($\tilde\Lambda=0$),
\begin{equation}
\frac{1}{\rho^2}\frac{d}{d\rho}\!\left[\rho(1-f)\right]=K(\rho),
\label{flat_einstein}
\end{equation}
one sees that the sole difference is the additive constant $\tilde\Lambda$ on the right-hand side of Eq.~(\ref{Einstein_first_order_AdS}). This is the mathematical expression of the fact, established in Sec.~\ref{sec:2}, that the cosmological constant modifies only the gravitational sector.

\paragraph{Decoupling of the mass equation.} We now make the key observation that allows the AdS mass function to be computed from the same equation as in the flat case. We define $f(\rho)$ through the 
generalized Schwarzschild-AdS ansatz
\begin{equation}
f(\rho)
=1-\frac{2m(\rho)}{\rho}-\frac{\tilde\Lambda}{3}\,\rho^2,
\label{f_ads_dimless}
\end{equation}
which explicitly separates the Misner--Sharp mass function $m(\rho)$ from the AdS background contribution $-(\tilde\Lambda/3)\rho^2$. Substituting Eq.~(\ref{f_ads_dimless}) into the left-hand side of 
Eq.~(\ref{Einstein_first_order_AdS}):
\begin{align}
\frac{1}{\rho^2}\frac{d}{d\rho}\!\left[\rho(1-f)\right]
&=
\frac{1}{\rho^2}\frac{d}{d\rho}\!\left[
\frac{2m(\rho)}{\rho}\cdot\rho
+\frac{\tilde\Lambda}{3}\rho^3
\right]\nonumber\\
&=
\frac{2}{\rho^2}\frac{dm}{d\rho}
+\tilde\Lambda.
\label{lhs_expand}
\end{align}
Equating with the right-hand side of Eq.~(\ref{Einstein_first_order_AdS}) gives
\begin{equation}
\frac{2}{\rho^2}\frac{dm}{d\rho}
+\tilde\Lambda
=K(\rho)+\tilde\Lambda,
\end{equation}
from which $\tilde\Lambda$ cancels \emph{identically}, yielding
\begin{equation}
\frac{dm}{d\rho}=\frac{\rho^2}{2}K(\rho).
\label{mass_eq}
\end{equation}
This is exactly the same first-order equation that governs the mass function in the asymptotically flat geometry. The cancellation of $\tilde\Lambda$ is not accidental: it is a direct consequence of having placed the cosmological contribution inside the definition~(\ref{f_ads_dimless}), i.e., of having correctly separated the physical mass $m(\rho)$ from the AdS background geometry. The result means that the entire electromagnetic content, encoded in $K(\rho)$, sources the mass function in an identical way regardless of whether the spacetime is asymptotically flat or asymptotically AdS.

\subsection{Parametric mass function and AdS lapse function}

Since Eq.~(\ref{mass_eq}) is identical to its flat-space counterpart, its solution is also the same. Integration along the parametric curve $\rho=\rho(y)$ gives
\begin{equation}
\rho(y)
=\left(\frac{2q^2}{y\,(1+y^{n-1})^3}\right)^{1/4},
\label{rhoy}
\end{equation}
and the mass function, obtained by integrating $dm/d\rho=({\rho^2}/{2})K(\rho)$ using the substitution $d\rho=(d\rho/dy)dy$, reads
\begin{widetext}
\begin{align}
m(y)
&=
\frac{q^{3/2}}{2^{1/4}\,15(n-1)}
\Bigg[
\frac{
n(17n-49)
+\bigl(10+n(27n-101)\bigr)y^{n-1}
-32n\,y^{2(n-1)}
}{
4n(1+y^{n-1})^{9/4}
}\,y^{1/4}\notag\\&
+\frac{8}{y^{(n-2)/4}}\,
{}_2F_1\!\!\left(
\tfrac{1}{4},\,
\tfrac{n-2}{4(n-1)},\,
\tfrac{5n-6}{4(n-1)},\,
-y^{1-n}
\right)
\Bigg]
+m_{\rm BH}-\bar m_{\rm field},
\label{my}
\end{align}
\end{widetext}
where ${}_2F_1$ is the Gauss hypergeometric function, and the field-mass normalization constant is
\begin{equation}
\bar m_{\rm field}
=
\frac{2^{3/4}q^{3/2}}{3}
\frac{
\Gamma\!\left(\dfrac{4n-3}{4(n-1)}\right)
\Gamma\!\left(\dfrac{5n-6}{4(n-1)}\right)
}{
\Gamma\!\left(\dfrac{9}{4}\right)
}.
\label{mfield}
\end{equation}

Several remarks on the structure of $m(y)$ are in order.

\emph{(i) Integration constant.} The free parameter $m_{\rm BH}$ is the physical ADM mass of the black hole, defined as the value of the total mass function at spatial infinity. The constant $\bar m_{\rm field}$ is the 
electromagnetic field-energy contribution evaluated at infinity; it is subtracted to ensure that $m_{\rm BH}$ represents only the gravitational (black-hole) mass, not the total field energy. This decomposition mirrors 
the analogous splitting in the Reissner--Nordstr\"{o}m case, where the gravitational mass and the electromagnetic field energy are separately identifiable.

\emph{(ii) Hypergeometric function.} The appearance of ${}_2F_1$ in Eq.~(\ref{my}) reflects the non-elementary nature of the integral $\int d\rho\,\rho^2 K(\rho(y))$ for generic $n$. For $n=2$, the hypergeometric function reduces to an algebraic expression, and the mass function simplifies to
\begin{equation}
m(y)\big|_{n=2}
=
m_{\rm BH}
+\frac{q^{3/2}}{2^{1/4}\cdot 15}
\frac{(-30-84y-64y^2)\,y^{1/4}}{8\,(1+y)^{9/4}},
\label{m_n2}
\end{equation}
which is the primary case used in the numerical analysis of this paper.

\emph{(iii) Asymptotic limits.} As $y\to 0$ (i.e.\ $\rho\to\infty$, the weak-field region), the electromagnetic correction to $m(y)$ vanishes and $m(y)\to m_{\rm BH}$, so the lapse function approaches the 
Schwarzschild-AdS form $f\to 1-2m_{\rm BH}/\rho+\rho^2/\tilde L^2$, as expected. As $y\to\infty$ (i.e.\ $\rho\to 0$, the strong-field region), the mass function approaches a finite limit governed by $\bar m_{\rm field}$, which prevents the usual Reissner-Nordstr\"om singularity from being sourced by a $1/r^2$ electric field; instead, the singularity structure at $r=0$ is modified by the PINLED nonlinear coupling, though the curvature singularity is not entirely resolved for the $Y^n$ model.

\subsection{Full AdS lapse function and global structure}

Combining Eqs.~(\ref{f_ads_dimless}), (\ref{rhoy}), and (\ref{my}), the complete dimensionless lapse function of the PINLED AdS black hole in parametric form is
\begin{equation}
f_{\rm AdS}(y)
=1-\frac{2m(y)}{\rho(y)}-\frac{\tilde\Lambda}{3}\,\rho^2(y),
\label{fAdS_y}
\end{equation}
Since $\tilde\Lambda<0$, the last term is positive and dominates for large $\rho$, driving $f_{\rm AdS}\to+\infty$ as $\rho\to\infty$. It is therefore convenient to introduce the dimensionless AdS radius $\tilde L^2=-3/\tilde\Lambda>0$ and rewrite the lapse as
\begin{equation}
f_{\rm AdS}(y)
=1-\frac{2m(y)}{\rho(y)}+\frac{\rho^2(y)}{\tilde L^2}.
\label{fAdS_L}
\end{equation}
The three terms on the right-hand side have a transparent physical interpretation: the first term is the flat-space contribution from the topology of the sphere; the second term encodes the gravitational attraction 
sourced by the mass function $m(y)$, which includes both the black-hole mass $m_{\rm BH}$ and the electromagnetic field energy; and the third term is the anti-de Sitter curvature contribution, which acts as a confining gravitational potential and ensures that the spacetime is globally AdS rather than 
asymptotically flat.

Equation~(\ref{fAdS_L}) is the main result of this section: the \emph{consistent AdS completion} of the PINLED $Y^n$ black hole. Its derivation confirms that the procedure of adding $\Lambda$ directly to the action, rather than modifying the metric function by hand, yields an internally consistent solution in which all field equations, both electromagnetic and gravitational, are simultaneously satisfied.

\subsection{Comparison with the RN-AdS limit}

In the limit $\gamma\to 0$ (equivalently $y^{n-1}\to 0$, i.e.\ weak PINLED coupling), the constitutive relation gives $P_{\mu\nu}\to F_{\mu\nu}$, the energy density reduces to $K(y)\to y/4$, and the parametric relations simplify to $\rho(y)\to (2q^2/y)^{1/4}$, so that $y\to 2q^2/\rho^4$ and
\begin{equation}
m(y)\big|_{\gamma\to 0}
=m_{\rm BH}+\int^\infty_\rho \frac{\rho'^2}{2}\cdot\frac{q^2}{2\rho'^4}\,d\rho'
=m_{\rm BH}+\frac{q^2}{4\rho},
\end{equation}
giving the mass function of the Reissner--Nordstr\"{o}m solution. The lapse function then becomes
\begin{equation}
f_{\rm AdS}\big|_{\gamma\to 0}
=1-\frac{2m_{\rm BH}}{\rho}+\frac{q^2}{\rho^2}+\frac{\rho^2}{\tilde L^2},
\end{equation}
which is precisely the Reissner--Nordstr\"{o}m-AdS (RN-AdS) lapse function~\cite{Chamblin:1999tk}, confirming that the PINLED AdS geometry reduces to the standard RN-AdS black hole in the Maxwell limit. The nonlinear PINLED corrections, therefore, represent a controlled deformation of the RN-AdS family, parametrized by the coupling $\gamma$ (or equivalently by the scale $\ell$ through Eq.~(\ref{gamma_fixed})) and the nonlinearity index $n$.

\section{Horizons and Hawking temperature}\label{sec:4}

\subsection{Horizon structure}

The event horizon is defined by the largest root of
\begin{equation}
f_{\rm AdS}(\rho_h)=0,
\label{horizon_cond2}
\end{equation}
which, using Eq.~(\ref{f_ads_dimless}), takes the explicit form
\begin{equation}
1-\frac{2m(\rho_h)}{\rho_h}-\frac{\tilde\Lambda}{3}\rho_h^2=0.
\label{horizon_explicit}
\end{equation}
This is an implicit equation for $\rho_h$ because $m(\rho)$ is itself a nontrivial function of the radial coordinate determined by the mass equation~(\ref{mass_eq}).

Because $\tilde\Lambda<0$, the AdS term $-(\tilde\Lambda/3)\rho^2>0$ contributes positively to the lapse function at large $\rho$, ensuring that $f_{\rm AdS}(\rho)\to+\infty$ as $\rho\to\infty$. For small $\rho$, the electromagnetic repulsion encoded in $m(\rho)$ competes with both the gravitational attraction and the AdS curvature. The number of positive roots of Eq.~(\ref{horizon_cond2}), and hence the horizon structure, depends sensitively on the three dimensionless parameters $\tilde{M}$, $q$, and $\tilde{P}$ (or equivalently $\tilde\Lambda$), as well as on the nonlinearity index $n$.

\begin{figure}[tbp]
  \centering
  \includegraphics[width=0.9\columnwidth]{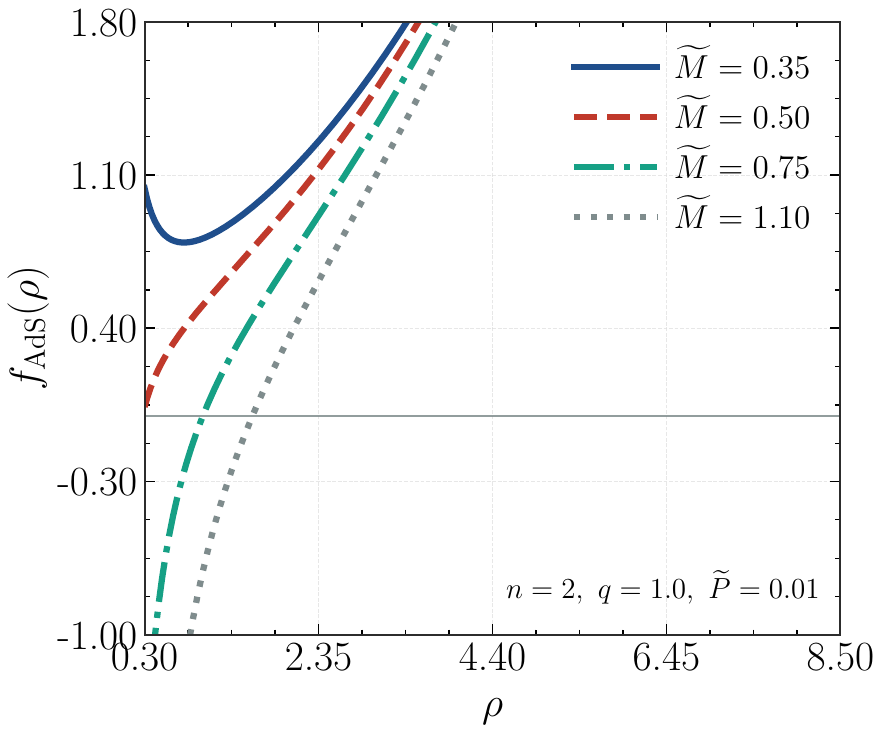}
  \caption{Lapse function $f_{\rm AdS}(\rho)$ for the PINLED AdS black hole with $n=2$, $q=1$, $\tilde P=0.01$, and four values of the dimensionless ADM mass $\widetilde M$. From bottom to top near the minimum, the curves correspond to no horizon (sub-extremal), extremal, two horizons, and a single large horizon.}
  \label{fig:lapse}
\end{figure}

Figure~\ref{fig:lapse} displays $f_{\rm AdS}(\rho)$ for $n=2$, $q=1$, $\tilde P=0.01$, and four representative values of the ADM mass $\tilde M$. Three qualitatively distinct regimes are apparent:
\begin{itemize}
  \item \emph{Sub-extremal} ($\tilde M=0.35$): the lapse function is strictly positive everywhere above a minimum radius, indicating the absence of a horizon. The geometry describes a naked electromagnetic source.
  \item \emph{Near-extremal} ($\tilde M=0.50$): $f_{\rm AdS}$ develops a local minimum that barely touches zero, corresponding to a degenerate (extremal) horizon.
  \item \emph{Two-horizon} ($\tilde M=0.75$): two distinct roots are present, yielding a Cauchy horizon $\rho_-$ and an event horizon $\rho_+$. This is the generic black-hole regime.
  \item \emph{Large-mass} ($\tilde M=1.10$): the inner Cauchy horizon shrinks, and the metric behaves much like the Schwarzschild-AdS case, with a single dominant event horizon.
\end{itemize}
These features are qualitatively similar to the Reissner--Nordstr\"{o}m-AdS geometry~\cite{Chamblin:1999tk}, but with the PINLED source encoding nonlinear corrections at the scale set by $\ell$ and $q$.

\subsection{Hawking temperature}

The Hawking temperature is obtained from the surface gravity at the event horizon,
\begin{equation}
T_H=\frac{\hbar\,\kappa_{\rm sg}}{2\pi k_B}
=\frac{1}{4\pi\ell}\,f'_{\rm AdS}(\rho_h),
\label{TH_def}
\end{equation}
where $f'_{\rm AdS}=df_{\rm AdS}/d\rho$. Using the first-order Einstein equation~(\ref{Einstein_first_order_AdS}) together with the horizon condition~(\ref{horizon_cond2}), the derivative evaluates to
\begin{equation}
f'_{\rm AdS}(\rho_h)
=\frac{1}{\rho_h}
-\rho_h K(\rho_h)
-\tilde\Lambda\rho_h,
\label{fderiv_h}
\end{equation}
so that the dimensionless temperature $\widetilde T\equiv\ell T_H$ is
\begin{equation}
\widetilde T
=\frac{1}{4\pi}\left[
\frac{1}{\rho_h}
-\rho_h K(y_h)
+8\pi\tilde P\,\rho_h
\right],
\label{temperature_final}
\end{equation}
where we used $-\tilde\Lambda=8\pi\tilde P$ and $K(y_h)$ denotes the dimensionless energy density evaluated at the horizon via the parametric relation $\rho_h=\rho(y_h)$.

Equation~(\ref{temperature_final}) has a transparent physical interpretation. The first term, $1/(4\pi\rho_h)$, is the inverse-radius contribution that dominates for small black holes. The second term, $-\rho_h K(y_h)/(4\pi)$, encodes the electromagnetic energy density at the horizon and is a purely nonlinear electrodynamics effect: it is absent in the RN-AdS case and, for the PINLED model, suppresses the temperature relative to the Schwarzschild-AdS baseline. The third term, $2\tilde P\rho_h$, is the AdS pressure contribution and causes the temperature to grow linearly for large $\rho_h$, a hallmark of asymptotically AdS geometries.

\subsection{Numerical results: temperature curves}

The interplay of these three contributions produces a rich $\widetilde T$-$\rho_h$ diagram. Since Eq.~(\ref{temperature_final}) involves only the parametric functions $\rho(y_h)$ and $K(y_h)$, the temperature can be computed without reference to the mass function $m(y)$: it is sufficient to sweep $y_h\in(0,\infty)$ and record the pair $(\rho_h,\widetilde T)$.

\begin{figure}[tbp]
  \centering
  \includegraphics[width=0.9\columnwidth]{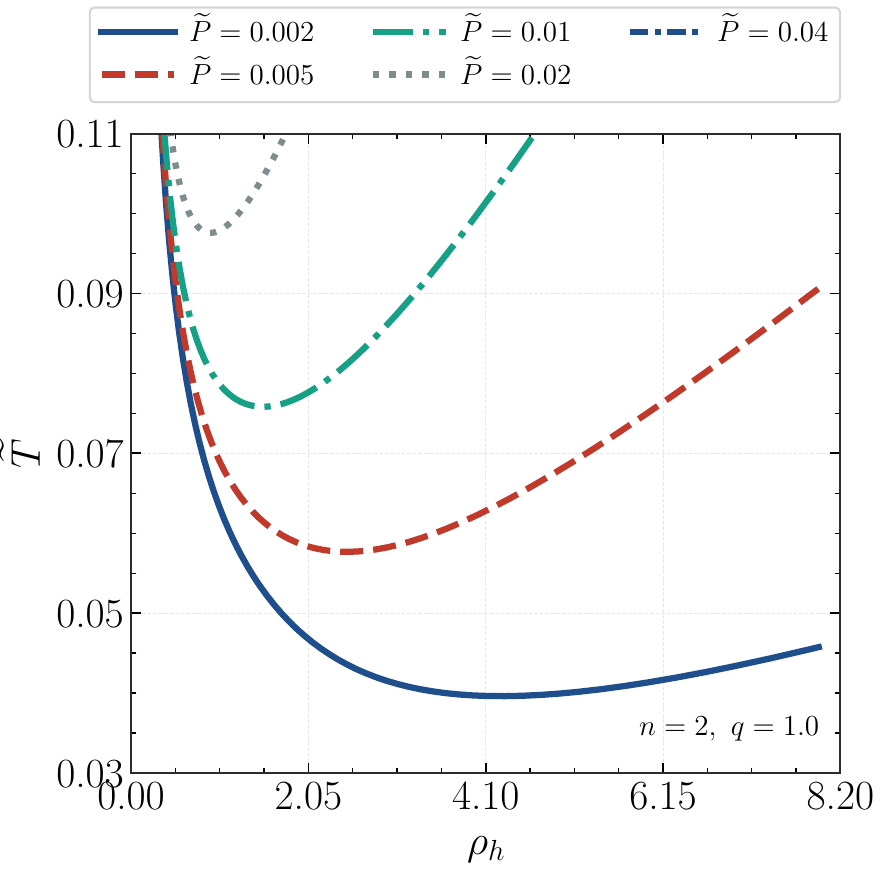}
  \caption{Dimensionless Hawking temperature $\widetilde T$ as a function of the horizon radius $\rho_h$ for $n=2$, $q=1$, and five values of $\tilde P$. Each curve has a minimum $\widetilde T_{\rm min}$ that separates the thermally unstable small-black-hole branch (left) from the stable large-black-hole branch (right). Increasing $\tilde P$ shifts the minimum to higher $\rho_h$ and lower $\widetilde T$.}
  \label{fig:T_pressure}
\end{figure}

\paragraph{Effect of pressure.}
Figure~\ref{fig:T_pressure} shows $\widetilde T$ as a function of $\rho_h$ for $n=2$, $q=1$, and five values of $\tilde P$. Each curve exhibits a global minimum $\widetilde T_{\rm min}$ at a radius $\rho_h^*$. For $\widetilde T>\widetilde T_{\rm min}$, there are two branches: a small-black-hole (SBH) branch with $\rho_h<\rho_h^*$ and a large-black-hole (LBH) branch with $\rho_h>\rho_h^*$. As $\tilde P$ increases, the AdS pressure term dominates at larger $\rho_h$, which causes the LBH branch to rise more steeply and pushes $\widetilde T_{\rm min}$ to smaller values and larger radii. This behavior is qualitatively analogous to the RN-AdS system~\cite{Kubiznak:2012wp}, but the PINLED nonlinear coupling shifts the minimum position and modifies the slope of the SBH branch.

\begin{figure}[tbp]
  \centering
  \includegraphics[width=0.9\columnwidth]{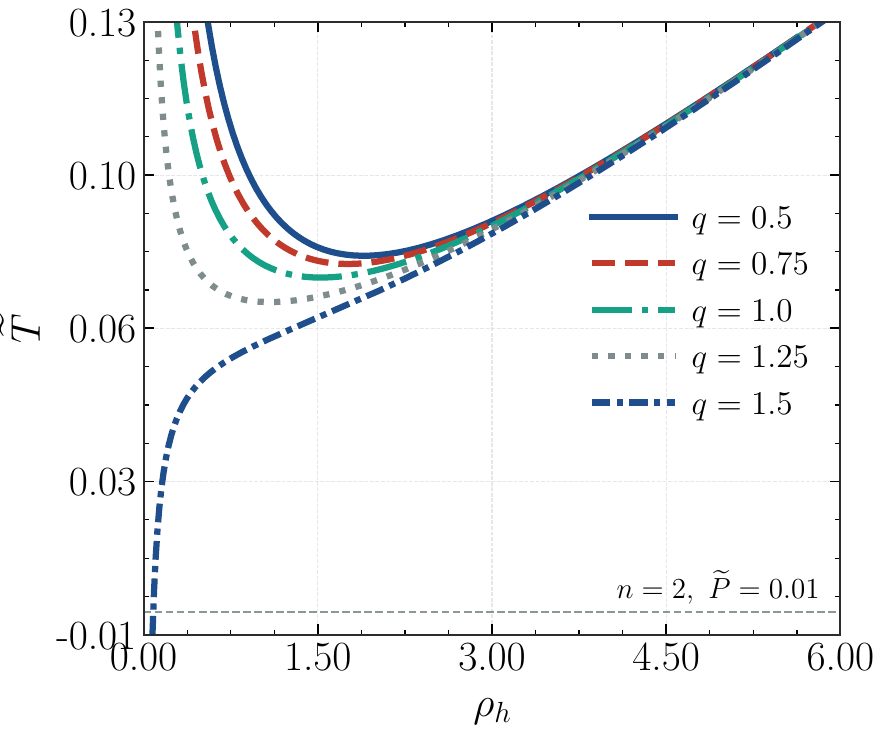}
  \caption{Dimensionless Hawking temperature $\widetilde T$ vs.\ $\rho_h$ for $n=2$, $\tilde P=0.01$, and five values of $q$. Higher charge suppresses the temperature and shifts the minimum toward larger $\rho_h$. The temperature is bounded below by $\widetilde T=0$, which defines the extremal radius for each $q$.}
  \label{fig:T_charge}
\end{figure}

\paragraph{Effect of charge.}
Figure~\ref{fig:T_charge} displays the temperature curves for fixed $n=2$, $\tilde P=0.01$, and varying $q$. Increasing the charge $q$ enlarges the electromagnetic contribution $\rho_h K(y_h)$ at any given $\rho_h$, which suppresses the temperature on the SBH branch. As a result, the minimum $\widetilde T_{\rm min}$ decreases with increasing $q$, and the extremal limit $\widetilde T\to 0$ is approached at finite $\rho_h$. At very large charge, the SBH branch entirely disappears (the temperature curve lies below zero for small $\rho_h$), signaling the absence of a thermodynamically accessible small black hole, a PINLED analog of the near-extremal RN-AdS behaviour~\cite{Chamblin:1999hg}.

\begin{figure}[tbp]
  \centering
  \includegraphics[width=0.9\columnwidth]{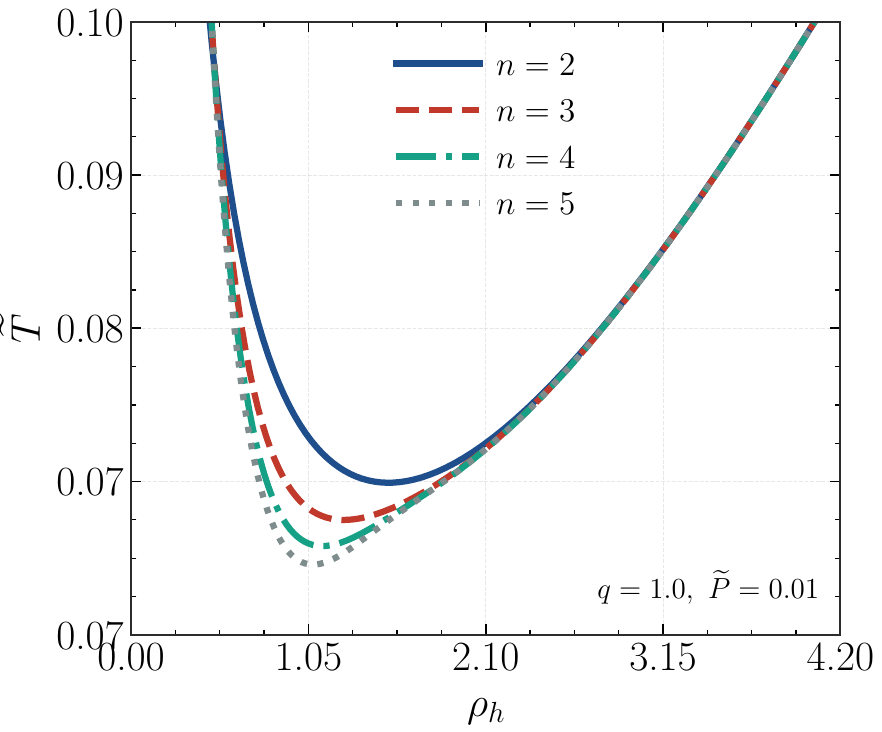}
  \caption{Dimensionless temperature $\widetilde T$ vs.\ $\rho_h$ for $q=1$, $\tilde P=0.01$, and four values of the PINLED nonlinearity index $n=2,3,4,5$. The curves merge on the large-black-hole branch and separate on the small-black-hole branch, where larger $n$ implies stronger electromagnetic suppression.}
  \label{fig:T_n}
\end{figure}

\paragraph{Effect of the nonlinearity index $n$.}
Figure~\ref{fig:T_n} compares the temperature curves for $n=2,3,4,5$ at fixed $q=1$ and $\tilde P=0.01$. The energy density function $K(y)=(y/4)+[(3n-2)/(4n)]y^n$ grows faster with $y$ for larger $n$, which means the electromagnetic suppression term $\rho_h K(y_h)$ is more pronounced at small horizon radii. Consequently, as $n$ increases, the SBH branch is more strongly suppressed, and the minimum temperature is shifted to a larger radius. On the LBH branch, all curves converge because the $y_h\to 0$ limit makes $K\to 0$, restoring the universal pressure-dominated behavior $\widetilde T\approx 2\tilde P\rho_h$.

\paragraph{Minimum temperature and extremal analysis.}

\begin{figure}[tbhp]
  \centering
  \includegraphics[width=0.9\columnwidth]{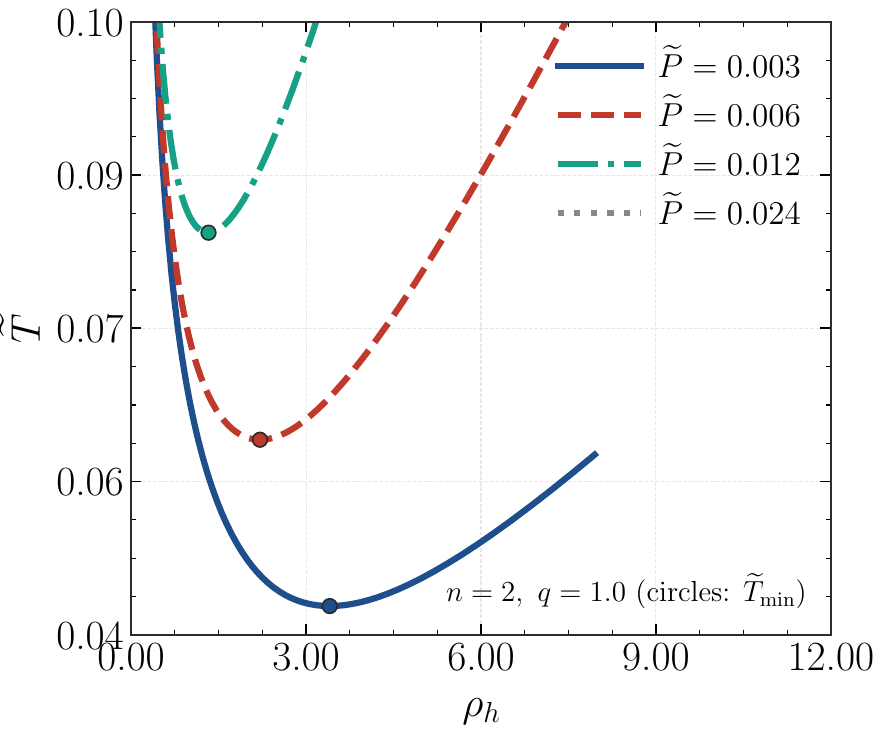}
  \caption{Dimensionless Hawking temperature $\widetilde T$ vs.\ $\rho_h$ for $n=2$, $q=1$, and four pressures. Filled circles mark the minimum $\widetilde T_{\rm min}$ at $\rho_h^*$, which corresponds to the divergence of $C_P$ and the onset of a second-order phase transition.}
  \label{fig:T_Pmin}
\end{figure}

Figure~\ref{fig:T_Pmin} shows, for $n=2$ and $q=1$, the temperature curves for four pressures with the minimum $\widetilde T_{\rm min}$ explicitly marked (filled circles). The minimum is determined by the stationarity condition $d\widetilde T/d\rho_h=0$, which gives
\begin{equation}
-\frac{1}{\rho_h^2}-K(\rho_h)-\rho_h K'(\rho_h)+8\pi\tilde P=0.
\end{equation}

This same combination reappears as the denominator of the heat capacity at constant pressure in Sec.~\ref{sec:5}; therefore, the temperature minimum also marks the divergence of $C_P$. The minimum temperature thus coincides with a second-order phase transition point in the extended phase space. As $\tilde P$ increases, $\widetilde T_{\rm min}$ decreases and $\rho_h^*$ shifts to larger values, indicating that the thermodynamically stable large-black-hole branch dominates at higher pressures. In the limit $\tilde P\to 0$, the minimum temperature approaches the extremal value.

\subsection{Large-radius limit and AdS asymptotics}

For large event horizons ($\rho_h\gg 1$, i.e., $y_h\to 0$), the energy density $K(y_h)\to 0$ and Eq.~(\ref{temperature_final}) reduces to
\begin{equation}
\widetilde T
\approx
\frac{1}{4\pi\rho_h}+2\tilde P\rho_h
\quad (\rho_h\gg 1).
\label{T_large}
\end{equation}
This is the standard Schwarzschild-AdS temperature in dimensionless units~\cite{Hawking:1982dh}, confirming that the nonlinear electrodynamics becomes negligible at large scales and the geometry reduces to the expected AdS behavior. The temperature then has a global minimum at $\rho_h^*=1/\sqrt{8\pi\tilde P}$, with $\widetilde T_{\rm min}^{\rm large}=\sqrt{\tilde P/(2\pi)}$. For finite $q$ and $n$, the PINLED corrections shift this minimum to larger $\rho_h^*$ and lower $\widetilde T_{\rm min}$, as confirmed by the numerical curves.

\subsection{Small-radius limit and near-extremal behavior}

For small event horizons ($\rho_h\to 0$, $y_h\to\infty$), the energy density grows as $K(y_h)\sim[(3n-2)/(4n)]y_h^n$, and the parametric relation gives $\rho_h\sim (2q^2)^{1/4}/y_h$ for large $y_h$. Therefore,
\begin{equation}
\rho_h K(y_h)
\sim
\frac{3n-2}{4n}\,(2q^2)^{1/4}\,y_h^{n-1}
\to\infty
\quad (y_h\to\infty).
\label{rhoK_limit}
\end{equation}
This divergence of the electromagnetic term suppresses the temperature faster than the $1/\rho_h$ divergence of the gravitational term in Eq.~(\ref{temperature_final}), ultimately driving $\widetilde T$ through zero. The zero of the temperature defines the extremal horizon radius $\rho_h^{\rm ext}$, below which no thermodynamically consistent black hole exists. All parameter combinations shown in the figures possess such an extremal point, although for large $\tilde P$ it lies at very small $\rho_h$ and is outside the displayed range.

\section{Extended phase-space thermodynamics}\label{sec:5}

In the extended phase-space framework, the cosmological constant is promoted to a thermodynamic variable by identifying it with the pressure~\cite{Kastor:2009wy,Dolan:2010ha,Dolan:2011xt}
\begin{equation}
P=-\frac{\Lambda}{8\pi G}
=\frac{3}{8\pi\ell^2\tilde{L}^2},
\qquad
\tilde P\equiv P\ell^2=-\frac{\tilde\Lambda}{8\pi},
\label{pressure_def}
\end{equation}
so that $-\tilde\Lambda=8\pi\tilde P>0$. In this interpretation, the black-hole mass $M=\ell\,\tilde M$ plays the role of enthalpy. At a fixed charge, the dimensionless entropy and thermodynamic volume are
\begin{equation}
\tilde S=\pi\rho_h^2,
\qquad
\tilde V=\frac{4\pi}{3}\rho_h^3,
\label{volume}
\end{equation}
and the extended first law becomes
\begin{equation}
d\tilde M=\tilde T\,d\tilde S+\tilde V\,d\tilde P.
\label{first_law}
\end{equation}
The Gibbs free energy
\begin{equation}
\tilde G=\tilde M-\tilde T\,\tilde S
=m_{\rm BH}-\pi\rho_h^2\tilde T
\label{Gibbs_def}
\end{equation}
then governs the phase structure. Below, we derive and analyze these quantities for the PINLED $Y^n$ model.

\subsection{Mass function on the horizon and Gibbs free energy}

The horizon condition $f_{\rm AdS}(\rho_h)=0$ determines the mass function at the horizon,
\begin{equation}
m(\rho_h)=\frac{\rho_h}{2}\left(1+\frac{\rho_h^2}{\tilde L^2}\right)
=\frac{\rho_h}{2}\left(1+\frac{8\pi\tilde P\rho_h^2}{3}\right).
\label{m_at_horizon}
\end{equation}
Since the PINLED mass function satisfies $m(y)=m_{\rm BH}+\delta m(y)$, where $\delta m(y)\leq 0$ encodes the electromagnetic field contribution (which is negative because the field energy outside the horizon is subtracted from the ADM mass), the physical mass parameter is
\begin{align}
&m_{\rm BH}=\frac{\rho_h}{2}\left(1+\frac{8\pi\tilde P\rho_h^2}{3}\right)-\delta m(y_h),
\\
&\delta m(y_h)=C\cdot A(y_h),
\label{mBH_formula}
\end{align}
where, for $n=2$,
\begin{equation}
C\cdot A(y)
=
\frac{q^{3/2}}{2^{1/4}\cdot 15}
\cdot
\frac{(-30-84y-64y^2)\,y^{1/4}}{8\,(1+y)^{9/4}}\leq 0.
\label{CAdelta}
\end{equation}
Substituting into Eq.~(\ref{Gibbs_def}) and using Eq.~(\ref{temperature_final}), the Gibbs free energy takes the closed parametric form
\begin{equation}
\tilde G(y_h)
=
\frac{\rho_h}{4}
+\left(\frac{4\pi}{3}-2\right)\tilde P\rho_h^3
+\frac{\rho_h^3 K(y_h)}{4}
-\delta m(y_h),
\label{Gibbs_parametric}
\end{equation}
which, together with $\tilde T(y_h)$ from Eq.~(\ref{temperature_final}), yields the $\tilde G$-$\tilde T$ diagram by sweeping $y_h\in(0,\infty)$.

\subsection{Equation of state}

The equation of state is obtained by solving Eq.~(\ref{temperature_final}) for $\tilde P$:
\begin{equation}
\tilde P
=
\frac{\tilde T}{2\rho_h}
-\frac{1}{8\pi\rho_h^2}
+\frac{K(y_h)}{8\pi}.
\label{eos_explicit}
\end{equation}
Introducing the specific volume $v=2\rho_h$ (the analog of the molar volume in the Van der Waals system), this becomes
\begin{equation}
\tilde P
=
\frac{\tilde T}{v}
-\frac{1}{2\pi v^2}
+\frac{K(y(v/2))}{8\pi}.
\label{eos_v_explicit}
\end{equation}
The first two terms reproduce the Schwarzschild-AdS equation of state in the large-$v$ limit where $K\to 0$. The third term, proportional to the PINLED energy density evaluated at the horizon, encodes the nonlinear electrodynamics correction.

\begin{figure}[tbp]
  \centering
  \includegraphics[width=0.9\columnwidth]{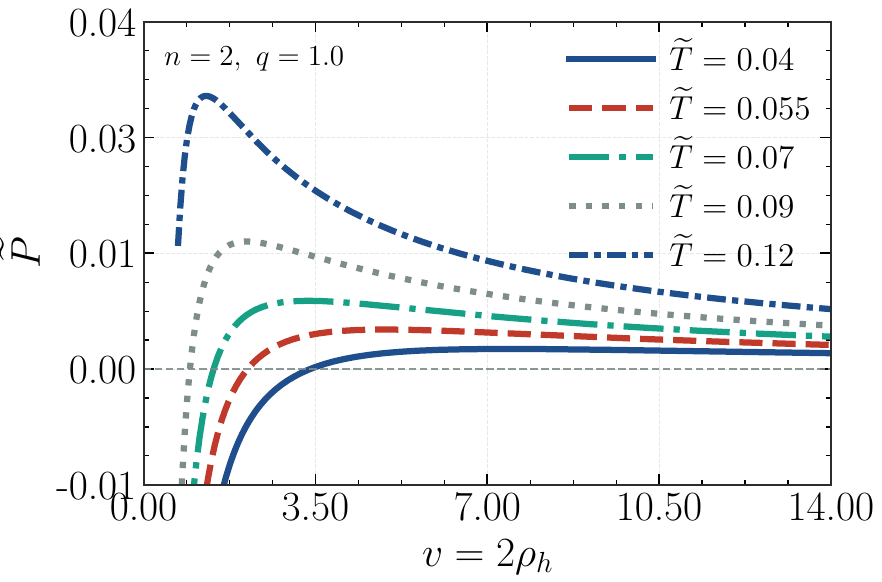}
  \caption{Equation-of-state isotherms $\tilde P$ vs.\ $v=2\rho_h$ for the PINLED AdS black hole with $n=2$, $q=1$, and five values of $\widetilde{T}$. All curves are monotone, indicating the absence of a Van der Waals critical point for this parameter set. The dashed horizontal line marks $\tilde P=0$.}
  \label{fig:eos}
\end{figure}

Figure~\ref{fig:eos} shows the $\tilde P$-$v$ isotherms for $n=2$, $q=1$, and five values of $\tilde T$. Remarkably, all isotherms remain smooth and do not develop the inflection-point structure characteristic of a Van der Waals critical point. For fixed $\tilde T$, $\tilde P$ rises as the horizon shrinks, reaches a maximum at intermediate $v$, and then decreases toward zero for large $v$. This behavior should be contrasted with the Reissner--Nordstr\"{o}m-AdS case, where the charge term $\propto v^{-4}$ creates a double root and a genuine VdW critical point~\cite{Kubiznak:2012wp}. The PINLED nonlinear coupling produces a softer electromagnetic correction ($\sim v^{-2}$ for large $v$) that is insufficient to generate a VdW critical point for $n=2$.

\subsection{Phase structure: Hawking-Page transition for n=2}

The absence of a VdW critical point in the $\tilde P$-$v$ plane for $n=2$ implies that the phase structure is governed instead by a \emph{Hawking-Page} (HP) transition~\cite{Hawking:1982dh} between the large black hole and thermal AdS.

\begin{figure}[tbp]
  \centering
  \includegraphics[width=0.9\columnwidth]{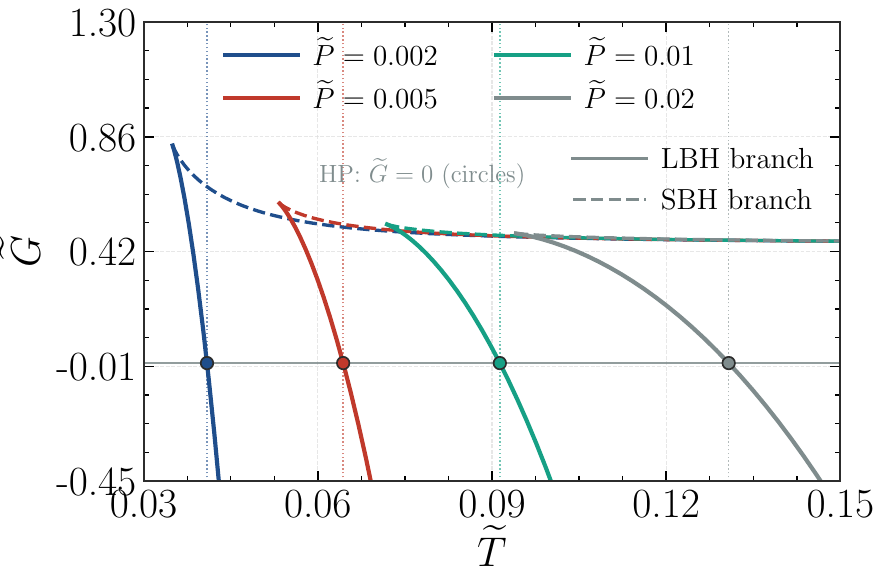}
  \caption{Gibbs free energy $\widetilde G$ as a function of temperature $\widetilde T$ for the PINLED AdS black hole with $n=2$, $q=1$, and four pressures $\tilde P$. Solid lines: LBH branch ($C_P>0$); dashed lines: SBH branch ($C_P<0$). Filled circles mark the Hawking-Page transition $\widetilde G=0$. Vertical dotted lines indicate the corresponding $\widetilde T_{\rm HP}$.}
  \label{fig:GT_n2}
\end{figure}

Figure~\ref{fig:GT_n2} shows $\tilde G$ as a function of $\tilde T$ for $n=2$, $q=1$, and four pressures. Each curve has a cusp at $(\tilde T_{\min},\tilde G_{\max})$, from which two branches emerge:
\begin{itemize}
\item \emph{Large black hole (LBH) branch} (solid lines): the stable branch with $C_P>0$; $\tilde G$ decreases rapidly from $\tilde G_{\max}$ as $\tilde T$ increases.
\item \emph{Small black hole (SBH) branch} (dashed lines): the locally unstable branch with $C_P<0$; $\tilde G>\tilde G_{\rm LBH}$ at every temperature, so the SBH is always thermodynamically subdominant relative to the LBH.
\end{itemize}
The LBH branch crosses $\tilde G=0$ at the Hawking-Page temperature $\tilde T_{\rm HP}$ (filled circles), marking a first-order phase transition between the large black hole and thermal AdS. For $\tilde T<\tilde T_{\rm HP}$, the thermal AdS background ($\tilde G=0$) is globally preferred; for $\tilde T>\tilde T_{\rm HP}$, the LBH is the stable phase.

\subsection{Emergence of the Van der Waals transition for n=3}

\begin{figure}[tbp]
  \centering
  \includegraphics[width=0.9\columnwidth]{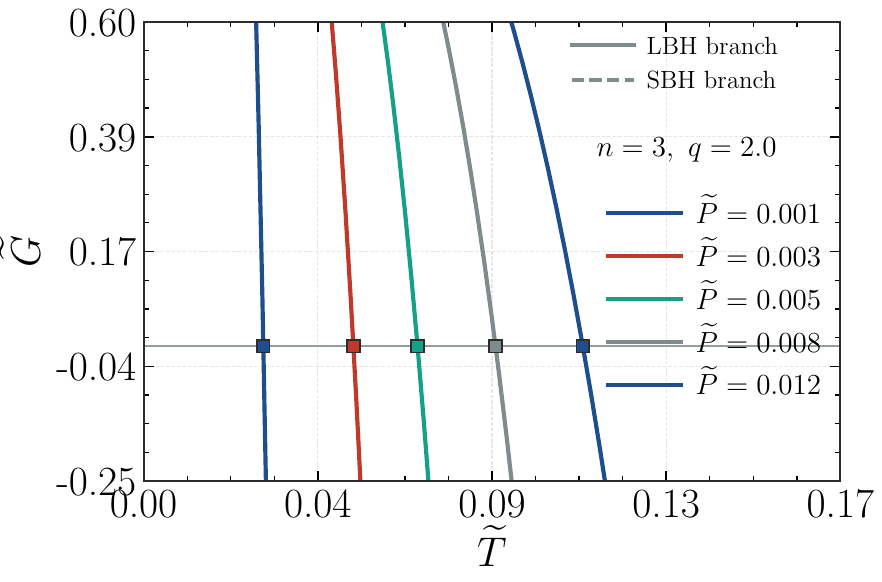}
  \caption{$\widetilde G$-$\widetilde T$ diagram for $n=3$, $q=2$, and five pressures. For small $\tilde P$, the SBH branch (dashed) descends below the LBH branch (solid), generating a swallowtail and a first-order SBH-LBH phase transition. Squares mark the Hawking-Page temperature $\widetilde T_{\rm HP}$.}
  \label{fig:GT_n3}
\end{figure}

The phase structure changes qualitatively for a higher nonlinearity index. Figure~\ref{fig:GT_n3} shows the $\tilde G$-$\tilde T$ diagram for $n=3$, $q=2$, and five pressures. In contrast to the $n=2$ case, for sufficiently small $\tilde P$ the $\tilde G$ curve develops a \emph{swallowtail} pattern: the SBH branch descends below the LBH branch at intermediate temperatures, creating a region where $\tilde G_{\rm SBH}<\tilde G_{\rm LBH}$ and, consequently, a first-order \emph{small-to-large} black hole phase transition. The transition temperature is determined by the crossing $\tilde G_{\rm SBH}(\tilde T^*)=\tilde G_{\rm LBH}(\tilde T^*)$, which is the black-hole analogue of the Van der Waals liquid-gas coexistence condition~\cite{Kubiznak:2012wp,Gunasekaran:2012dq}. As $\tilde P$ increases toward a critical value $\tilde P_c$, the swallowtail shrinks and the two branches merge, signaling a second-order critical point. Above $\tilde P_c$, the $\tilde G$-$\tilde T$ curve is single-valued (no swallowtail) and the phase transition is replaced by a Hawking-Page transition.

This result demonstrates that the PINLED model supports a richer phase structure than a single fixed nonlinearity index would suggest: the parameter $n$ controls the functional form of the electromagnetic energy density $K(y)$ and hence the effective charge contribution in the EoS. For $n=2$, the electromagnetic correction is too soft to generate a VdW critical point at the parameters considered; for $n\geq 3$ with sufficiently large charge, the critical point emerges and the full VdW phase structure is recovered.

\subsection{Heat capacity at constant pressure}

The heat capacity at constant pressure,
\begin{align}
C_P
&=
T_H\!\left(\frac{\partial S}{\partial T_H}\right)_P \notag\\&
=
2\pi\ell^2\rho_h
\frac{
1/\rho_h-\rho_h K+8\pi\tilde P\rho_h
}{
-1/\rho_h^2-K-\rho_h(\partial K/\partial\rho_h)+8\pi\tilde P
},
\label{Cp_explicit}
\end{align}
provides a local stability criterion: $C_P>0$ on the LBH branch and $C_P<0$ on the SBH branch. The divergence of $C_P$ occurs when the denominator in Eq.~(\ref{Cp_explicit}) vanishes, which coincides exactly with the temperature minimum condition $d\tilde T/d\rho_h=0$. At this point, the system transitions between locally stable and unstable behavior.

\begin{figure}[tbp]
  \centering
  \includegraphics[width=0.9\columnwidth]{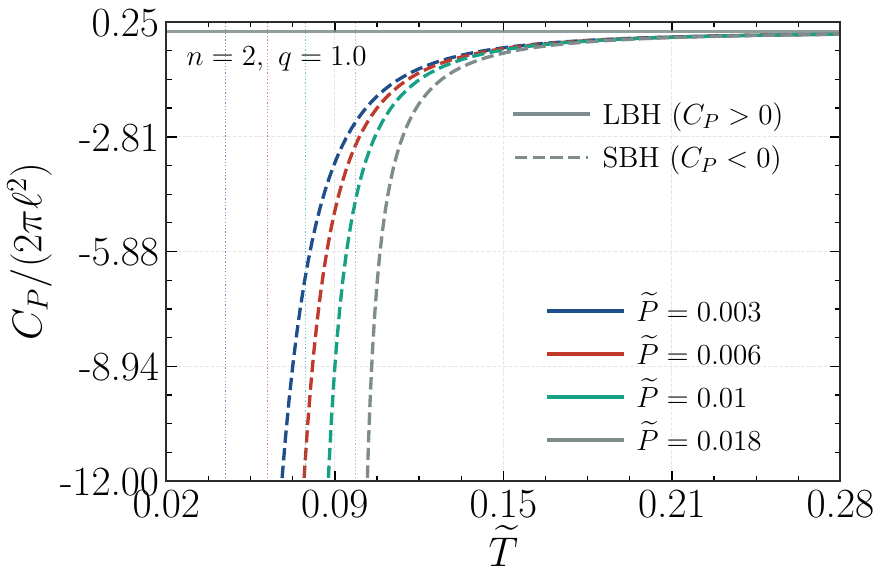}
  \caption{Heat capacity $C_P/(2\pi\ell^2)$ as a function of $\widetilde T$ for $n=2$, $q=1$, and four pressures. Solid (dashed) lines correspond to the LBH (SBH) branch with $C_P>0$ ($C_P<0$). Vertical dotted lines mark $\widetilde T_{\min}$, where $C_P$ diverges.}
  \label{fig:Cp}
\end{figure}

Figure~\ref{fig:Cp} shows $C_P/(2\pi\ell^2)$ as a function of $\tilde T$ for $n=2$, $q=1$, and four pressures. On the LBH branch (solid lines), $C_P$ is positive and increases with $\tilde T$, characteristic of thermally stable black holes. On the SBH branch (dashed lines), $C_P$ is negative, signaling local thermodynamic instability. Both branches diverge at $\tilde T=\tilde T_{\min}$ (indicated by vertical dotted lines), where the heat capacity changes sign and the two branches connect. The divergence shifts to higher $\tilde T$ as $\tilde P$ increases, consistent with the behavior of $\tilde T_{\min}(\tilde P)$ shown in Fig.~\ref{fig:THP}.

\subsection{Hawking-Page temperature and phase diagram}

\begin{figure}[tbp]
  \centering
  \includegraphics[width=0.9\columnwidth]{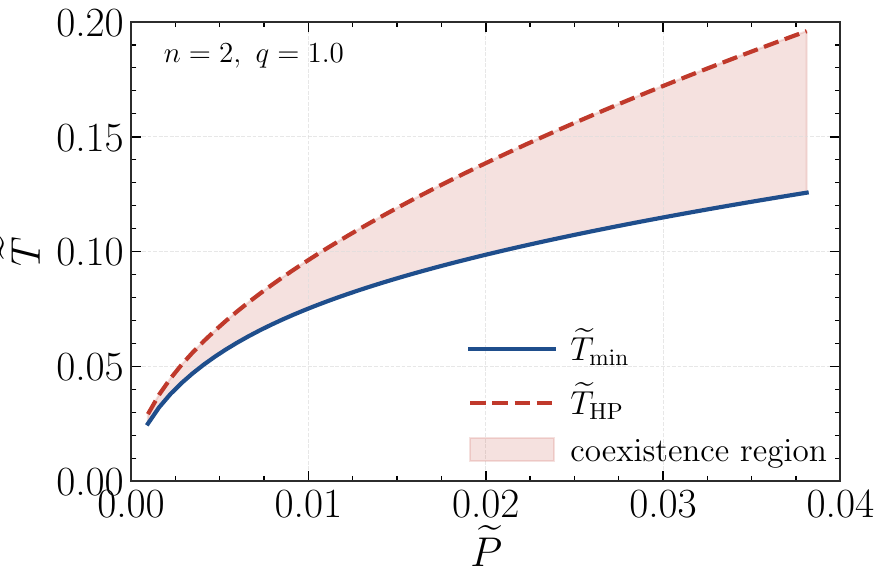}
  \caption{Phase diagram in the $\tilde P$-$\widetilde T$ plane for the PINLED AdS black hole with $n=2$, $q=1$. The solid (blue) curve is the minimum temperature $\widetilde T_{\min}$ below which no black holes exist. The dashed (red) curve is the Hawking-Page temperature $\widetilde T_{\rm HP}$ (first-order transition). The shaded band marks the region where black holes exist but thermal AdS is preferred.}
  \label{fig:THP}
\end{figure}

Figure~\ref{fig:THP} shows the minimum temperature $\tilde T_{\min}$ and the Hawking-Page temperature $\tilde T_{\rm HP}$ as functions of $\tilde P$ for $n=2$, $q=1$. Three thermodynamic regions are identified:
\begin{itemize}
\item $\tilde T<\tilde T_{\min}(\tilde P)$: no black hole solution exists; the system is in the thermal AdS phase.
\item $\tilde T_{\min}<\tilde T<\tilde T_{\rm HP}$: black holes exist (both SBH and LBH branches), but $\tilde G>0$; the thermal AdS background remains globally preferred (shaded region).
\item $\tilde T>\tilde T_{\rm HP}$: the LBH has $\tilde G<0$ and is the globally stable phase; the HP transition occurs at $\tilde T_{\rm HP}$.
\end{itemize}
Both $\tilde T_{\min}$ and $\tilde T_{\rm HP}$ increase with $\tilde P$, indicating that higher AdS pressure stabilizes the large black hole. The shaded coexistence region shrinks as $\tilde P$ increases, consistent with the LBH becoming thermodynamically accessible at progressively lower temperatures relative to $\tilde T_{\min}$.

\subsection{Summary of the thermodynamic phase structure}

The extended phase-space analysis of the PINLED $Y^n$ model reveals the following hierarchy:
\begin{enumerate}
\item For $n=2$, the electromagnetic correction $K(y)\sim y/4$ produces a softer $\tilde P$-$v$ isotherm than RN-AdS, and the system undergoes only a Hawking-Page first-order transition between thermal AdS and a stable large black hole. The SBH is locally unstable ($C_P<0$) and thermodynamically subdominant at all temperatures.
\item For $n\geq 3$ at sufficiently large charge $q$, the energy density $K(y)\sim [(3n-2)/(4n)]y^n$ grows faster with $y$, generating a stronger electromagnetic contribution at small horizon radii. This can produce a swallowtail in the $\tilde G$-$\tilde T$ diagram and a genuine VdW-like SBH-LBH first-order phase transition, qualitatively analogous to the RN-AdS case~\cite{Kubiznak:2012wp,Gunasekaran:2012dq}.
\end{enumerate}
Thus, the nonlinearity index $n$ plays a key role in determining the universality class of the black-hole phase transition: $n=2$ yields the Hawking-Page universality class, while $n\geq 3$ at large $q$ crosses over to the Van der Waals universality class. This sensitivity to the model index is a distinctive feature of the PINLED $Y^n$ family.

\section{Null and timelike geodesics}\label{sec:6}

We now turn to the geodesic structure of the PINLED AdS black hole. The analysis of null geodesics, which govern light propagation, yields the photon sphere and the shadow radius, both of which are directly observable. Timelike geodesics, which govern the motion of massive test particles, determine the circular-orbit structure and the innermost stable circular orbit (ISCO), which sets the inner edge of the accretion disk. Throughout this section, we work in the equatorial plane $\theta=\pi/2$.

\subsection{Geodesic equations and effective potentials}

The geodesic Lagrangian in the dimensionless equatorial plane is
\begin{equation}
\mathcal{L}_{\rm geo}
=
-g_{\mu\nu}\dot x^\mu\dot x^\nu
=
-f_{\rm AdS}(\rho)\,\dot t^2
+\frac{\dot\rho^2}{f_{\rm AdS}(\rho)}
+\rho^2\dot\phi^2,
\label{Lag2}
\end{equation}
where dots denote differentiation with respect to an affine parameter. The stationarity and axisymmetry of the metric yield two conserved quantities, the specific energy $E=f_{\rm AdS}(\rho)\,\dot t$ and the specific angular momentum $L=\rho^2\dot\phi$. Setting $\mathcal{L}_{\rm geo}=-\xi$ with $\xi=0$ (null) or $\xi=1$ (timelike), the radial equation takes the standard form
\begin{equation}
\dot\rho^2 + V_{\rm eff}(\rho) = E^2,
\qquad
V_{\rm eff}(\rho) = f_{\rm AdS}(\rho)\!\left(\frac{L^2}{\rho^2}+\xi\right).
\label{Veff2}
\end{equation}

\subsubsection{Null effective potential}

\begin{figure}[tbp]
  \centering
  \includegraphics[width=0.9\columnwidth]{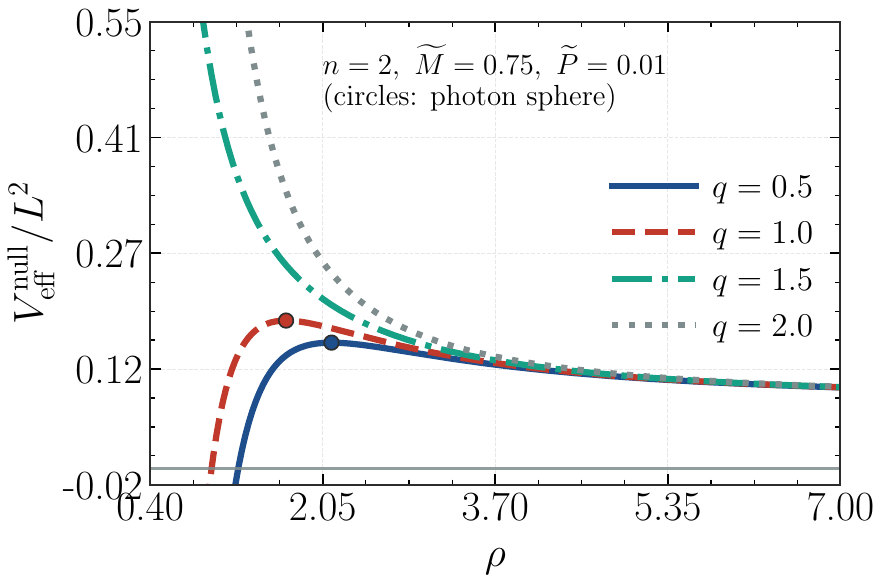}
  \caption{Null effective potential $V_{\rm eff}^{\rm null}/L^2$ as a function of $\rho$ for $n=2$, $\widetilde M=0.75$, $\widetilde P=0.01$, and four values of $q$. Filled circles mark the photon-sphere radius $\rho_{\rm ph}$. Increasing $q$ shifts $\rho_{\rm ph}$ outward and reduces the height of the potential barrier.}
  \label{fig:Veff_null}
\end{figure}

For $\xi=0$,
\begin{equation}
V_{\rm eff}^{\rm null}(\rho)=f_{\rm AdS}(\rho)\,\frac{L^2}{\rho^2}.
\label{Veff_null2}
\end{equation}
The maximum of $V_{\rm eff}^{\rm null}$ defines the photon sphere, which separates captured from scattered photon trajectories. Figure~\ref{fig:Veff_null} shows $V_{\rm eff}^{\rm null}/L^2$ for $n=2$, $\widetilde M=0.75$, $\widetilde P=0.01$, and four values of the charge $q$. Each potential exhibits a single maximum whose height, the capture cross-section, decreases as $q$ increases: a larger nonlinear charge lowers the potential barrier and reduces the capture cross-section, analogous to the effect of charge in the Reissner--Nordstr\"{o}m case but with a modified radial dependence. The photon-sphere radii (filled circles) shift outward with increasing $q$, reflecting the interplay between the electromagnetic repulsion and the AdS curvature.

\begin{figure}[tbp]
  \centering
  \includegraphics[width=0.9\columnwidth]{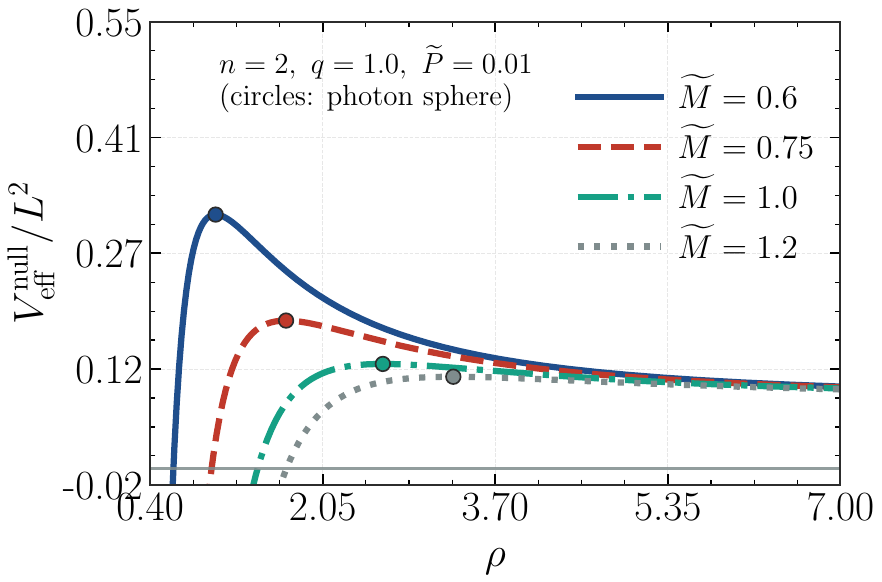}
  \caption{Null effective potential $V_{\rm eff}^{\rm null}/L^2$ for $n=2$, $q=1$, $\widetilde P=0.01$, and four values of $\widetilde M$. The photon sphere (filled circles) shifts outward, and the barrier grows taller with increasing mass.}
  \label{fig:Veff_null_mass}
\end{figure}

Figure~\ref{fig:Veff_null_mass} shows the same potential for fixed $q=1$, $\widetilde P=0.01$, and four values of the ADM mass $\widetilde M$. As $\widetilde M$ increases, the black hole grows and the photon-sphere shifts to larger $\rho_{\rm ph}$, while the height of the barrier increases monotonically. This is qualitatively consistent with the Schwarzschild-AdS limit, where $\rho_{\rm ph}\to 3\widetilde M$ for large masses, but modified by the PINLED electromagnetic field.

\subsubsection{Timelike effective potential}

\begin{figure}[tbp]
  \centering
  \includegraphics[width=0.9\columnwidth]{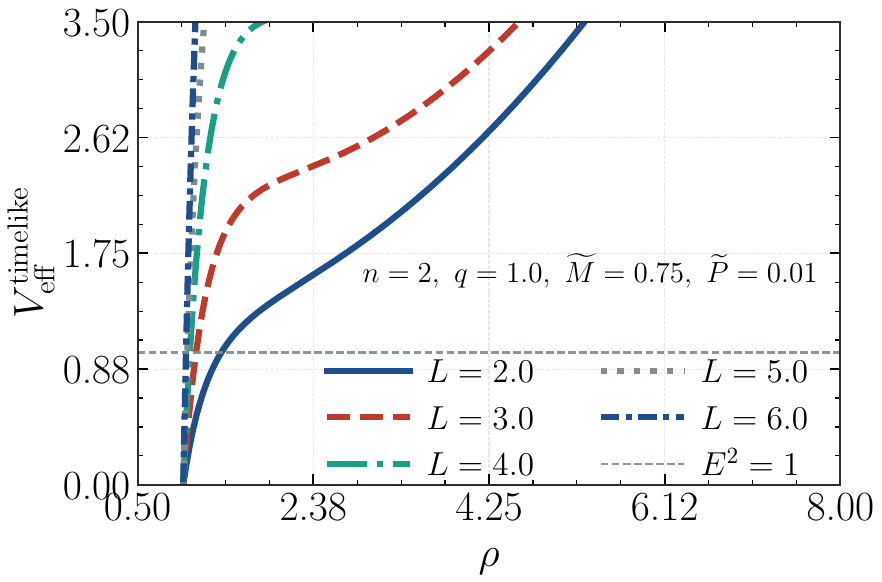}
  \caption{Timelike effective potential $V_{\rm eff}^{\rm timelike}$ for $n=2$, $q=1$, $\widetilde M=0.75$, $\widetilde P=0.01$, and five values of $L$. The dashed horizontal line indicates $E^2=1$ (rest energy). For $L\gtrsim 3$, a local minimum appears, allowing stable circular orbits.}
  \label{fig:Veff_time}
\end{figure}

For $\xi=1$,
\begin{equation}
V_{\rm eff}^{\rm timelike}(\rho)
=f_{\rm AdS}(\rho)\!\left(\frac{L^2}{\rho^2}+1\right).
\label{Veff_time2}
\end{equation}
Figure~\ref{fig:Veff_time} shows $V_{\rm eff}^{\rm timelike}$ for $n=2$, $q=1$, $\widetilde M=0.75$, $\widetilde P=0.01$, and five values of $L$. For small $L$ (e.g.\ $L=2$), the potential is monotonically increasing beyond the horizon; no local minimum exists, and no stable circular orbit is available. As $L$ increases, a local minimum develops, signaling the existence of stable circular orbits. The energy $E^2=1$ line (dashed horizontal) intersects the potential on the right side of the local maximum, producing bounded trajectories. The innermost such orbit, the ISCO, is defined by the simultaneous conditions $V'_{\rm eff}=V''_{\rm eff}=0$.

\subsection{Photon sphere}

The photon sphere is determined by the condition
\begin{equation}
\rho_{\rm ph}\,f'_{\rm AdS}(\rho_{\rm ph})=2\,f_{\rm AdS}(\rho_{\rm ph}),
\label{ps_cond}
\end{equation}
which can be rewritten using the first-order Einstein equation~(\ref{Einstein_first_order_AdS}) as
\begin{equation}
1-3f_{\rm AdS}(\rho_{\rm ph})
-\rho_{\rm ph}^2\,K(\rho_{\rm ph})
+8\pi\tilde P\,\rho_{\rm ph}^2
=0.
\label{ps_analytic}
\end{equation}
This is a transcendental equation for $\rho_{\rm ph}$ because $K$ depends on $\rho_{\rm ph}$ through the parametric relation $\rho_{\rm ph}=\rho(y_{\rm ph})$. 

\begin{figure*}[tbhp]
  \centering
  \includegraphics[scale=0.5]{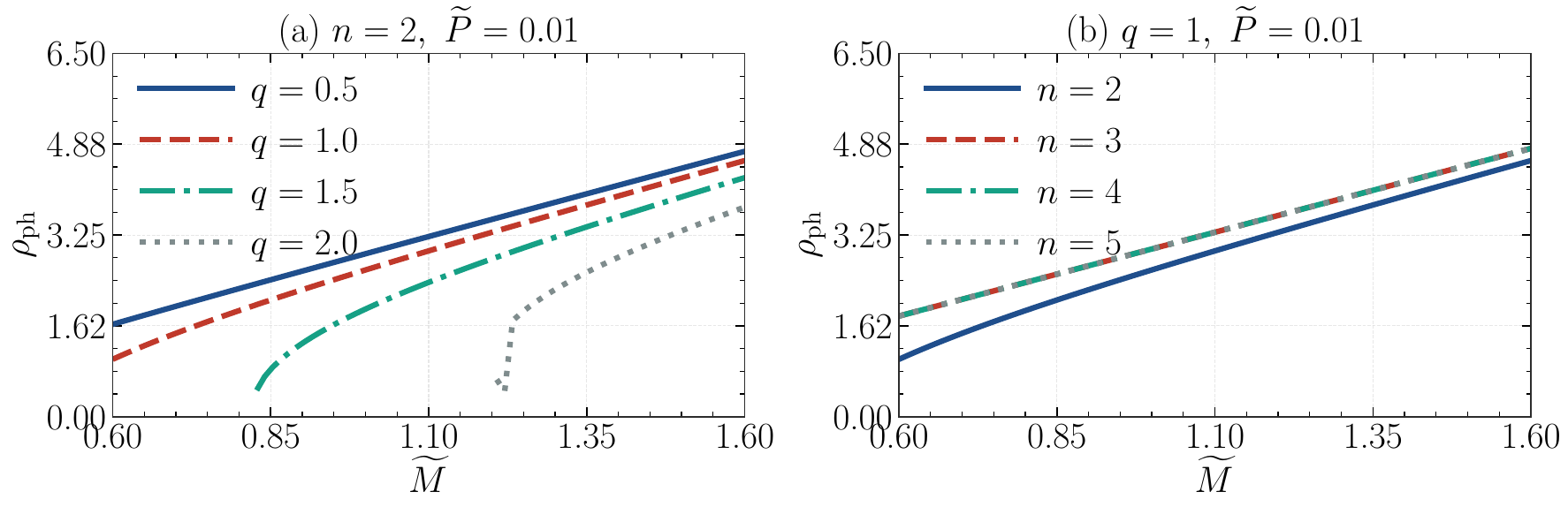}
  \caption{Photon-sphere radius $\rho_{\rm ph}$ as a function of the ADM mass $\widetilde M$ for $\widetilde P=0.01$. In (a) $n=2$, varying $q$. In (b) $q=1$, varying $n$. In both cases $\rho_{\rm ph}$ grows monotonically with $\widetilde M$; larger $q$ or $n$ shifts $\rho_{\rm ph}$ slightly inward.}
  \label{fig:photon_sphere}
\end{figure*}

Figure~\ref{fig:photon_sphere} shows $\rho_{\rm ph}$ as a function of $\widetilde M$ for two panels: (left) $n=2$ with varying $q$, and (right) $q=1$ with varying $n$. In both cases, $\rho_{\rm ph}$ grows monotonically with $\widetilde M$, as expected from the increasing gravitational potential. A larger charge $q$ or a larger nonlinearity index $n$ shifts the photon sphere slightly inward at fixed $\widetilde M$, reflecting the electromagnetic repulsion that effectively weakens the gravitational capture. For very small $\widetilde M$ (close to the minimum mass for black-hole existence), no photon sphere solution exists in the displayed range, consistent with the absence of an event horizon noted in Sec.~\ref{sec:4}.

\subsection{Shadow radius for a finite-distance observer}

In an asymptotically AdS spacetime, the natural observable is the shadow angular radius as seen by a static observer at finite radius $\rho_{\rm obs}$. The apparent shadow radius in the observer's image plane is~\cite{Perlick:2015vta,Pantig:2022whj}
\begin{equation}
\rho_s(\rho_{\rm obs})
=
\rho_{\rm ph}
\sqrt{\frac{f_{\rm AdS}(\rho_{\rm obs})}{f_{\rm AdS}(\rho_{\rm ph})}}.
\label{shadow_r}
\end{equation}
This expression generalizes the flat-space formula $\rho_s^{\rm flat}=\rho_{\rm ph}/\sqrt{f(\rho_{\rm ph})}$ by including the redshift factor $\sqrt{f_{\rm AdS}(\rho_{\rm obs})}$ at the observer's location. For large $\rho_{\rm obs}$, $f_{\rm AdS}(\rho_{\rm obs})\approx\rho_{\rm obs}^2/\tilde L^2$ and $\rho_s$ grows without bound, reflecting the fact that an observer at spatial infinity would see an infinitely large shadow in AdS, the shadow is a local observable at finite $\rho_{\rm obs}$.

\begin{figure*}[tbp]
  \centering
  \includegraphics[scale=0.47]{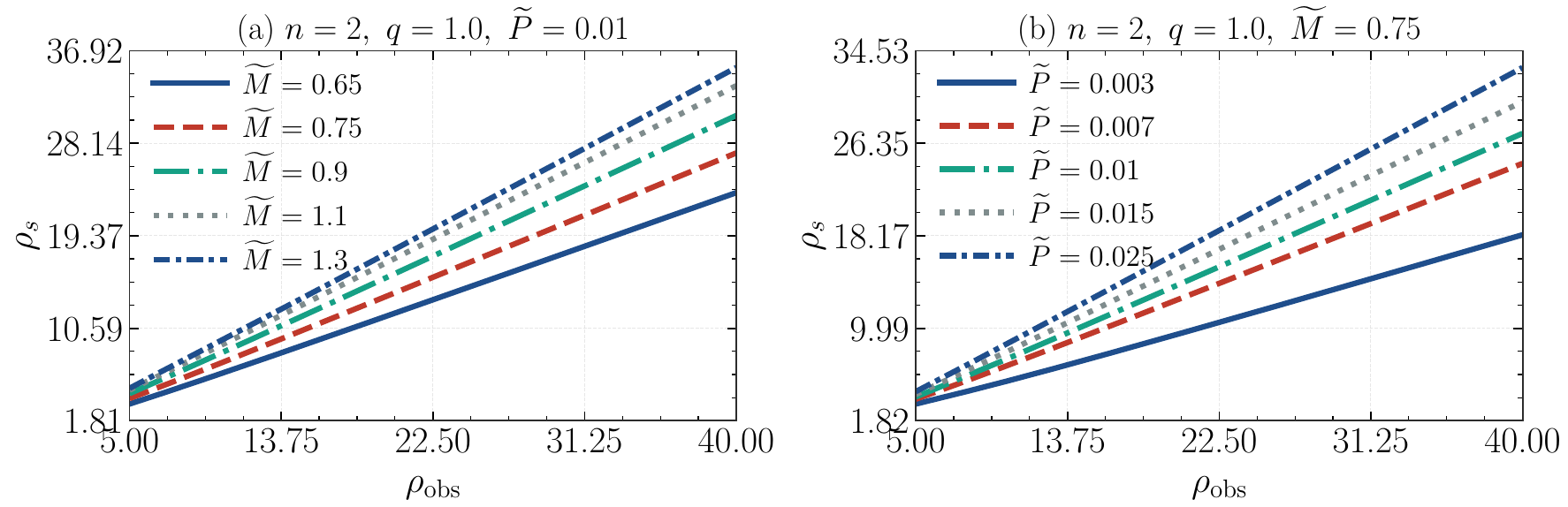}
  \caption{Shadow radius $\rho_s$ as a function of the observer position $\rho_{\rm obs}$ for $n=2$, $q=1$. In (a): varying $\widetilde M$ at $\widetilde P=0.01$. In (b): varying $\widetilde P$ at $\widetilde M=0.80$. In both cases, $\rho_s$ increases monotonically with $\rho_{\rm obs}$, reflecting the AdS magnification effect.}
  \label{fig:shadow}
\end{figure*}

Figure~\ref{fig:shadow} shows $\rho_s$ as a function of $\rho_{\rm obs}$ for (left) varying $\widetilde M$ at fixed $q=1$, $n=2$, $\widetilde P=0.01$, and (right) varying $\widetilde P$ at fixed $q=1$, $n=2$, $\widetilde M=0.80$. In both panels, $\rho_s$ increases monotonically with $\rho_{\rm obs}$, as the AdS potential amplifies the apparent size for more distant observers. A larger ADM mass $\widetilde M$ yields a larger photon sphere and hence a larger shadow at every $\rho_{\rm obs}$. A larger pressure $\widetilde P$ (smaller AdS radius $\tilde L$) compresses the geometry and shifts the shadow curve downward at fixed $\rho_{\rm obs}$, since the AdS curvature contributes to the lapse function both at the photon sphere and at the observer.

\begin{figure*}[tbhp]
  \centering
  \includegraphics[width=\textwidth]{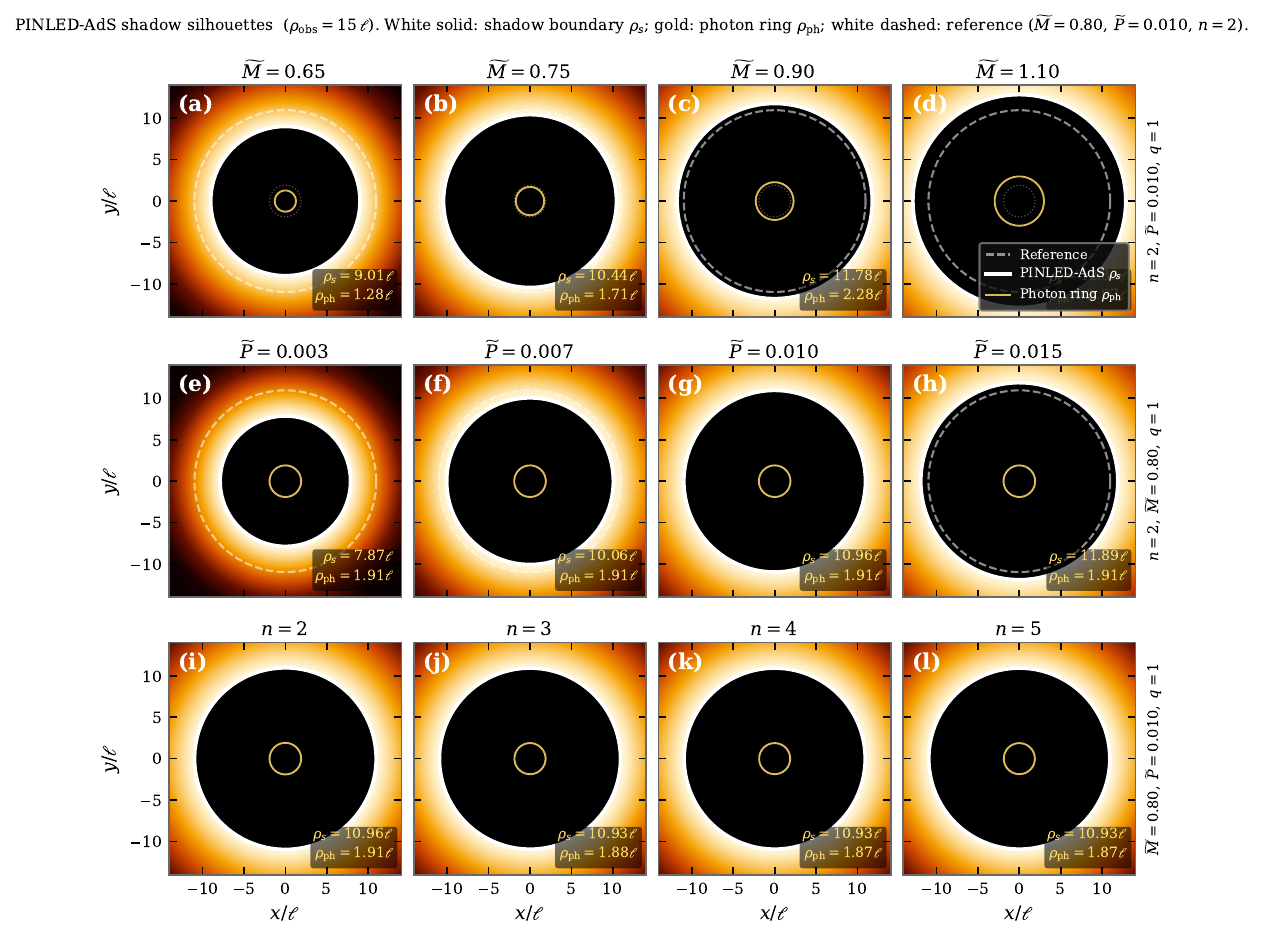}
  \caption{%
    Shadow silhouettes of the PINLED-AdS black hole as seen by a
    static observer located at $\rho_{\rm obs}=15\,\ell$, computed
    from Eq.~\eqref{shadow_r}.
    The background rendering mimics the thermal emission of a
    geometrically thin accretion disk; the solid white circle is the
    apparent shadow boundary $\rho_s$, the solid gold circle marks the
    photon-ring radius $\rho_{\rm ph}$, and the white dashed circle
    is the common reference silhouette
    $(\widetilde{M}=0.80,\,\widetilde{P}=0.010,\,n=2,\,q=1)$
    with $\rho_s^{\rm ref}=10.96\,\ell$ and
    $\rho_{\rm ph}^{\rm ref}=1.90\,\ell$.
    Top row~(a)\,--\,(d): $n=2$, $\widetilde{P}=0.010$, $q=1$,
    and $\widetilde{M}=0.65,\,0.75,\,0.90,\,1.10$.
    Both $\rho_s$ and $\rho_{\rm ph}$ grow monotonically with the ADM
    mass, consistent with the geodesic analysis of Sec.~\ref{sec:6}.
    Middle row~(e)\,--\,(h): $n=2$, $\widetilde{M}=0.80$,
    $q=1$, and $\widetilde{P}=0.003,\,0.007,\,0.010,\,0.015$.
    While $\rho_{\rm ph}$ is insensitive to pressure
    ($\rho_{\rm ph}\approx 1.90\,\ell$ throughout), $\rho_s$
    increases with $\widetilde{P}$: the larger AdS curvature amplifies
    the redshift factor $\sqrt{f_{\rm AdS}(\rho_{\rm obs})/
    f_{\rm AdS}(\rho_{\rm ph})}$ and hence the apparent size of the
    shadow for a finite-distance observer.
    Bottom row~(i)\,--\,(l): $\widetilde{M}=0.80$,
    $\widetilde{P}=0.010$, $q=1$, and $n=2,3,4,5$.
    Increasing the PINLED nonlinearity index produces only a mild
    inward shift of the photon sphere
    ($\rho_{\rm ph}=1.905,\,1.876,\,1.873,\,1.872\,\ell$ for
    $n=2,3,4,5$, respectively), leaving $\rho_s$ nearly unchanged.
    This saturation reflects the fact that, at fixed $\widetilde{M}$
    and $\widetilde{P}$, the dominant contribution to $\rho_s$ comes
    from the AdS redshift factor rather than from the PINLED
    electromagnetic correction to the photon-sphere location.
  }
  \label{fig:shadow_silhouettes}
\end{figure*}

Figure~\ref{fig:shadow_silhouettes} presents the apparent shadow
silhouettes of the PINLED-AdS black hole as recorded by a static
observer at the fiducial position $\rho_{\rm obs}=15\,\ell$, for
three independent parameter sweeps: the ADM mass $\widetilde{M}$
(top row), the AdS pressure $\widetilde{P}$ (middle row), and the
PINLED nonlinearity index $n$ (bottom row).
In each panel, the white solid circle represents the shadow boundary
$\rho_s$ computed from Eq.~\eqref{shadow_r}, the gold circle marks
the photon-ring radius $\rho_{\rm ph}$ determined by the photon-sphere
condition~\eqref{ps_cond}, and the white dashed circle is a common
reference silhouette evaluated at
$(\widetilde{M},\widetilde{P},n,q)=(0.80,\,0.010,\,2,\,1)$, which
yields $\rho_s^{\rm ref}=10.96\,\ell$ and
$\rho_{\rm ph}^{\rm ref}=1.90\,\ell$.

\paragraph{Dependence on the ADM mass.}
Panels~(a)--(d) fix $n=2$, $\widetilde{P}=0.010$, and $q=1$ while
varying $\widetilde{M}\in\{0.65,\,0.75,\,0.90,\,1.10\}$.
Both radii grow monotonically with mass: $\rho_{\rm ph}$ increases
from $1.27\,\ell$ to $2.97\,\ell$, and $\rho_s$ from
$9.01\,\ell$ to $12.85\,\ell$.
This behavior is qualitatively analogous to the Schwarzschild-AdS
limit, where $\rho_{\rm ph}\to 3\widetilde{M}$
for large masses, but the PINLED electromagnetic field shifts the
photon sphere to slightly smaller values relative to $3\widetilde{M}$,
as already noted in Fig.~\ref{fig:photon_sphere}.
The silhouettes in panels~(c) and~(d) are visibly larger than the
reference contour, while panel~(a) ($\widetilde{M}=0.65$) lies
noticeably inside it, furnishing a clear visual diagnostic of the
mass through shadow observations.

\paragraph{Dependence on the AdS pressure.}
Panels~(e)--(h) fix $n=2$, $\widetilde{M}=0.80$, $q=1$, and
vary $\widetilde{P}\in\{0.003,\,0.007,\,0.010,\,0.015\}$.
A striking feature emerges: the photon-sphere radius is entirely
insensitive to $\widetilde{P}$, remaining at
$\rho_{\rm ph}=1.90\,\ell$ across the full pressure range.
This decoupling is a direct consequence of the photon sphere
condition~\eqref{ps_analytic}, in which the terms proportional to
$8\pi\widetilde{P}\rho^2$ cancel against the AdS contribution to
$f_{\rm AdS}$ at the photon-sphere location for a fixed mass.
By contrast, the shadow radius $\rho_s$ does depend on pressure: it
rises from $7.87\,\ell$ at $\widetilde{P}=0.003$ to
$11.89\,\ell$ at $\widetilde{P}=0.015$, because the AdS curvature
amplifies the redshift factor
$\sqrt{f_{\rm AdS}(\rho_{\rm obs})/f_{\rm AdS}(\rho_{\rm ph})}$
in Eq.~\eqref{shadow_r} for a finite-distance observer.
This pressure-induced magnification has no analog in the
asymptotically flat case, and is a distinctive signature of the
AdS background.

\paragraph{Dependence on the nonlinearity index $n$.}
Panels~(i)--(l) fix $\widetilde{M}=0.80$, $\widetilde{P}=0.010$,
$q=1$, and vary $n\in\{2,3,4,5\}$.
The PINLED nonlinear coupling produces only a mild inward shift of the
photon sphere as $n$ increases:
$\rho_{\rm ph}=1.905,\,1.876,\,1.873,\,1.872\,\ell$
for $n=2,3,4,5$, respectively, corresponding to a total variation of
less than $2\%$.
The shadow radius $\rho_s$ is correspondingly stable at
$10.96,\,10.93,\,10.93,\,10.93\,\ell$, effectively saturating
already at $n=3$.
This saturation can be understood from the parametric relations: the
energy density $K(y)\sim[(3n-2)/(4n)]\,y^n$ grows with $n$ only in
the strong-field region $y\sim\mathcal{O}(1)$ (i.e., near the photon
sphere at $\rho\sim\ell$), while the AdS redshift factor that controls
$\rho_s$ is dominated by the weak-field region $y\to 0$
($\rho\gg\ell$), where $K\to 0$ independently of $n$.
As a result, the nonlinearity index primarily affects the near-horizon
thermodynamics and the ISCO (as discussed in Secs.~\ref{sec:5}
and~\ref{sec:6}), but leaves the observable shadow radius almost
unchanged for fixed $\widetilde{M}$ and $\widetilde{P}$.

\paragraph{Summary and observational implications.}
The three-parameter sweeps presented in
Fig.~\ref{fig:shadow_silhouettes} reveal a clear hierarchy of
sensitivity: the shadow radius $\rho_s$ is most strongly controlled
by the ADM mass $\widetilde{M}$ (variation of $\sim 43\%$ across the
displayed range), moderately sensitive to the AdS pressure
$\widetilde{P}$ (variation of $\sim 51\%$, attributable entirely to
the finite-observer redshift), and essentially insensitive to the
PINLED nonlinearity index $n$ (variation $<0.3\%$).
The photon-sphere radius $\rho_{\rm ph}$, on the other hand, responds
only to $\widetilde{M}$ and $n$, and is blind to $\widetilde{P}$.
This orthogonality between the sensitivities of $\rho_s$ and
$\rho_{\rm ph}$ to the model parameters suggests that a joint
measurement of both observables --- in principle accessible to
next-generation very-long-baseline interferometry instruments through
the separation of the lensing ring and the geometric shadow --- could
provide independent constraints on the mass and the AdS curvature
scale, while the nonlinearity index $n$ would need to be probed
through complementary thermodynamic or quasinormal-mode observables.

\subsection{Timelike circular orbits and the ISCO}

Circular timelike orbits satisfy $V'_{\rm eff}(\rho_c)=0$, yielding the specific angular momentum
\begin{equation}
L^2(\rho_c)
=
-\rho_c^2\,
\frac{1-f(\rho_c)-\rho_c^2 K(\rho_c)+8\pi\tilde P\rho_c^2}
{1-3f(\rho_c)-\rho_c^2 K(\rho_c)+8\pi\tilde P\rho_c^2}.
\label{L2_c}
\end{equation}
The ISCO corresponds to the inflection point of $L^2(\rho_c)$, i.e.\ to the simultaneous conditions $V'_{\rm eff}=0$ and $V''_{\rm eff}=0$, which in terms of $L^2(\rho)$ is equivalent to requiring $dL^2/d\rho_c=0$. This marginal-stability condition determines $\rho_{\rm ISCO}$.

\begin{figure*}[tbhp]
  \centering
  \includegraphics[scale=0.5]{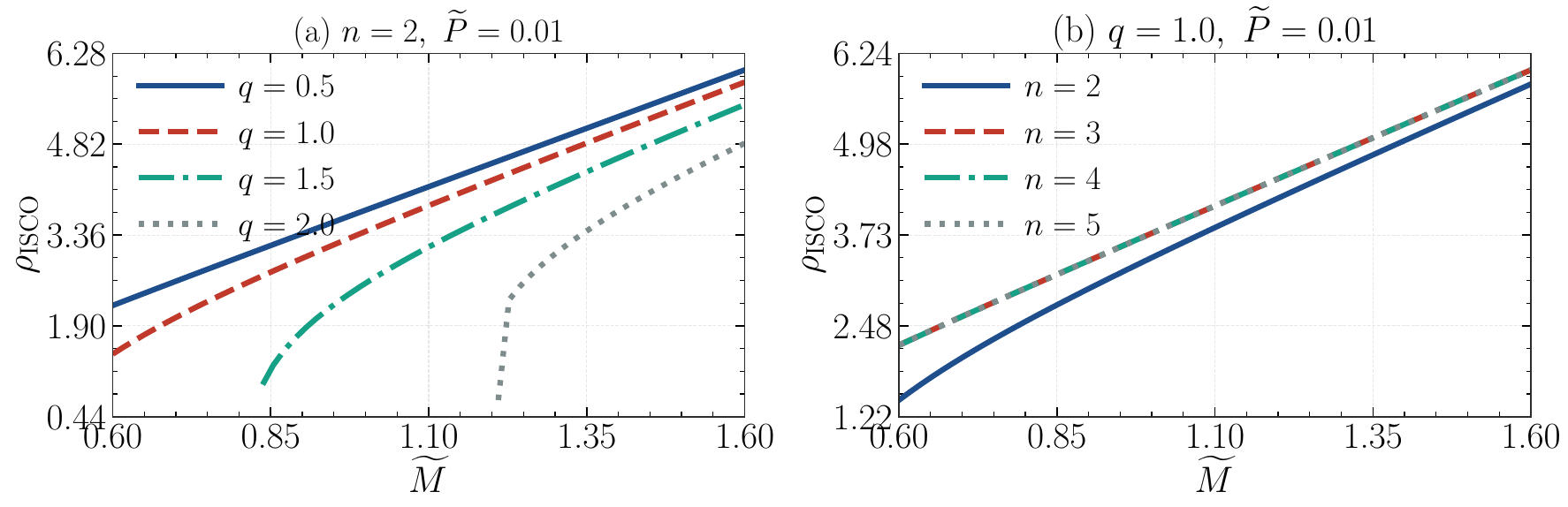}
  \caption{ISCO radius $\rho_{\rm ISCO}$ as a function of $\widetilde M$ for $\widetilde P=0.01$. In (a): $n=2$, varying $q$; in (b): $q=1$, varying $n$. In both panels, $\rho_{\rm ISCO}$ increases with $\widetilde M$, and a larger charge or nonlinearity index shifts the ISCO outward.}
  \label{fig:ISCO}
\end{figure*}

Figure~\ref{fig:ISCO} shows $\rho_{\rm ISCO}$ as a function of $\widetilde M$ for (left) $n=2$ with varying $q$, and (right) $q=1$ with varying $n$. The ISCO radius grows monotonically with $\widetilde M$ in all cases. Increasing $q$ at fixed $\widetilde M$ and $n$ shifts $\rho_{\rm ISCO}$ outward: the electromagnetic field provides an effective repulsive correction that destabilizes circular orbits at smaller radii, forcing the ISCO to larger values. This parallels the behavior of the RN solution, where the charge enlarges the ISCO relative to the Schwarzschild value $\rho_{\rm ISCO}^{\rm Schw}=6\widetilde M$. The dependence on $n$ is more subtle: a larger nonlinearity index produces a stronger short-range electromagnetic correction, which also shifts the ISCO outward at fixed $q$ and $\widetilde M$.

Qualitatively, the ISCO in the PINLED AdS geometry is bounded from below by the horizon radius and from above by the photon sphere. For the parameter values explored, $\rho_{\rm ph}<\rho_{\rm ISCO}<\rho_{\rm ISCO}^{\rm Schw}$ is not generally satisfied; the electromagnetic corrections can make $\rho_{\rm ISCO}>\rho_{\rm ISCO}^{\rm Schw}$, indicating a reduced radiative efficiency relative to the Schwarzschild-AdS baseline.

\subsection{Summary of geodesic observables}

Table~\ref{tab:geodesic} collects the key geodesic quantities for representative parameter choices, providing a compact reference for the orbital structure. The entries confirm the trends observed in the figures: both $\rho_{\rm ph}$ and $\rho_{\rm ISCO}$ are monotonically increasing functions of $\widetilde M$, $q$, and $n$, while the shadow radius $\rho_s$ at fixed $\rho_{\rm obs}=15$ also grows with $\widetilde M$ and is suppressed by larger $\widetilde P$.

\begin{table}[tbhp]
\centering
\caption{Geodesic observables for the PINLED AdS black hole with $n=2$, $q=1$, $\widetilde P=0.01$, and selected values of $\widetilde M$. Shadow radius evaluated at $\rho_{\rm obs}=15$.}
\label{tab:geodesic}
\begin{tabular}{cccc}
\hline\hline
$\widetilde M$ & $\rho_{\rm ph}$ & $\rho_{\rm ISCO}$ & $\rho_s(\rho_{\rm obs}=15)$ \\
\hline
$0.65$ & $1.269$ & $1.758$ & $8.975$ \\
$0.75$ & $1.700$ & $2.292$ & $10.41$ \\
$0.90$ & $2.271$ & $2.987$ & $11.77$ \\
$1.10$ & $2.966$ & $3.833$ & $12.84$ \\
$1.30$ & $3.625$ & $4.639$ & $13.49$ \\
\hline\hline
\end{tabular}
\end{table}

\FloatBarrier
\section{Conclusions}\label{sec:7}

We have constructed a consistent anti-de Sitter extension of the static and spherically symmetric PINLED $Y^n$ black hole by introducing the cosmological constant directly at the level of the Einstein--PINLED action. This procedure preserves the nonlinear electromagnetic sector and, with it, the original parametric relations among the electric field, the auxiliary tensor field, and the radial coordinate, while the gravitational sector acquires the standard AdS contribution in the lapse function. A key outcome of the derivation is that the mass equation remains the same as in the asymptotically flat case, so the AdS solution emerges as a genuine completion of the original model in the same gauge and parametrization rather than as an ad hoc deformation of the metric.

The thermodynamic analysis shows that this AdS completion exhibits rich, well-structured phase behavior. The horizon equation and the Hawking temperature preserve the nontrivial interplay among the gravitational term, the PINLED energy density, and the AdS pressure, yielding the familiar small and large-black-hole branches, along with an extremal limit. In the extended phase space, the $n=2$ sector is governed by a Hawking--Page transition between thermal AdS and a stable large black hole, whereas higher nonlinearity can generate a richer Van der Waals-like behavior when the charge is sufficiently large. The equation of state, Gibbs free energy, heat capacity, and phase diagram all consistently indicate that the nonlinearity index acts as an effective control parameter of the thermodynamic universality class. 

The geodesic analysis complements this thermodynamic picture by showing how the AdS background and the nonlinear electromagnetic sector jointly modify the effective potentials, the photon sphere, the shadow radius for a static observer at finite distance, and the innermost stable circular orbit. The numerical trends collected throughout the paper indicate that the mass, charge, AdS pressure, and nonlinearity index leave clear signatures on both orbital stability and optical observables. Altogether, the results presented here establish a self-consistent AdS realization of the PINLED $Y^n$ family and provide a solid starting point for future studies of quasinormal modes, accretion properties, shadow phenomenology, and broader comparisons with other nonlinear electrodynamics models.

\section*{Acknowledgments}

This work was partially supported by the Brazilian agencies CAPES, CNPq, and FAPEMA. EOS acknowledges the support from grants CNPq/306308/2022-3,
FAPEMA/UNIVERSAL-06395/22, and CAPES/Code 001. F.A. acknowledges the Inter University Center for Astronomy and Astrophysics (IUCAA), Pune, India, for granting visiting associateship. J.A.A.S.R acknowledges partial financial support from UESB through Grant AuxPPI (Edital No. 267/2024), as well as from FAPESB--CNPq/Produtividade under Grant No. 12243/2025 (TOB-BOL2798/2025).

\section*{Data Availability Statement}
There are no new data associated with this article.

\section*{Code/Software}
No code/software were developed in this article.

\end{document}